\documentclass[12pt]{iopart}
\usepackage{setspace}
\usepackage{mathptmx}
\usepackage[perpage]{footmisc}
\usepackage{lipsum}

\usepackage{amsmath}
\usepackage{graphicx}
\usepackage{cite}
\usepackage[usenames,dvipsnames,svgnames,table]{xcolor}
\usepackage{hyperref}
\hypersetup{
    colorlinks,
    linkcolor={red!50!black},
    citecolor={blue!50!black},
    urlcolor={blue!80!black}
}

\begin{document}
\newcommand{\nomarkfootnote}[1]{
\begingroup 
\renewcommand{\thefootnote}{\relax}%
\footnotetext{#1} 
\endgroup }
\onehalfspacing

\title{Plasmonic excitations in graphene layers}

\author{Pablo Mart\'in-Luna$^{1,*}$, Alexandre Bonatto$^2$, Cristian Bontoiu$^{3,4}$, Bifeng Lei$^{3,4}$, Guoxing Xia$^{5,4}$, Javier Resta-L\'opez$^{6,*}$
}

\address{$^1$ Instituto de F\'isica Corpuscular (IFIC), Universitat de Val\`encia - Consejo Superior de Investigaciones Cient\'ificas, 46980 Paterna, Spain}

\address{$^2$ Graduate Program in Information Technology and Healthcare Management, and the Beam Physics Group, Federal University of Health Sciences of Porto Alegre, Porto Alegre, RS, 90050-170, Brazil}

\address{$^3$ Department of Physics, The University of Liverpool, Liverpool L69 3BX, United Kingdom}

\address{$^4$ The Cockcroft Institute, Sci-Tech Daresbury, Warrington WA4 4AD, United Kingdom}

\address{$^5$ Department of Physics and Astronomy, The University of Manchester, Manchester M13 9PL, United Kingdom}

\address{$^6$ Instituto de Ciencia de los Materiales (ICMUV), Universidad de Valencia, 46071 Valencia, Spain}

\footnotetext{\, Corresponding authors: \href{mailto:pablo.martin@uv.es}{pablo.martin@uv.es} (Pablo Mart\'in-Luna), \href{mailto:javier2.resta@uv.es}{javier2.resta@uv.es} (Javier Resta-L\'opez)}

\vspace{10pt}
\begin{indented}
\item[]Keywords: graphene layers, wakefield, hydrodynamic model, plasmons \\

\item[]\today
\end{indented}

\begin{abstract}
The interaction of fast charged particles with graphene layers can generate electromagnetic modes. This wake effect has been recently proposed for short-wavelength, high-gradient particle acceleration and for obtaining brilliant radiation sources. In this study, the excitation of wakefields produced by a point-like charged particle moving parallel to a multilayer graphene array (which may be supported by an insulated substrate) is studied using the linearized hydrodynamic theory. General expressions for the excited longitudinal and transverse wakefields have been derived. The dependencies of the wakefields on the positions of the layers and the substrate, the velocity and the surface density have been extensively analyzed. This study provides a deeper understanding of the physical phenomena underlying plasmonic excitations in graphene layers, paving the way for potential applications of these structures in particle acceleration, nanotechnology and materials science.

\end{abstract}

\maketitle


\section{Introduction}\label{Introduction}
Graphene, a single sheet of carbon atoms forming a hexagonal lattice, is intimately related to carbon nanotubes (CNTs), which can be considered as a rolled sheet of graphene, and fullerenes, which are spherical molecules derived from graphene by introducing pentagons. 
Since its discovery in 2004 by Novoselov and
others \cite{Novoselov2004_graphene}, graphene has attracted great interest for applications in different areas (e.g. electronics, optics, THz technology, energy storage, biotechnology or medical science) due to its exceptional electrical, thermo-mechanical and optical properties \cite{LUO2013_plasmons_in_graphene_applications}. 
For this reason, the vibrational properties \cite{Yan2008_vibrational_graphene_PhysRevB.77.125401, wang2009vibrational_graphene, cooper2012experimental_vibrational_graphene} and the electronic properties \cite{Nilsson2006_electronic_properties_graphene_PhysRevLett.97.266801, Novoselov2007_electronic_properties_graphene, CastroNeto2009_electronic_properties_graphene_RevModPhys.81.109} in graphene layers have been widely studied in both theoretical and experimental aspects.

Carbon-based nanostructures are currently under active investigation for wakefield acceleration \cite{Bonatto2023_effective_plasma_POP, Bontoiu2023_catapult, Martin-Luna2023_ExcitationWakefieldsSWCNT_NJP, Martin-Luna2024_DWCNTs_ResultsInPhysics}. These structures present potential as efficient acceleration systems, since their wider channels in two dimensions allow to overcome the angstrom-size limitation of natural crystals—initially proposed by T. Tajima and others \cite{TajimaCavenago1987_XrayAccelerator_PhysRevLett.59.1440, chen1987solid, chen1997crystal} in the 1980s and 1990s—as solid-state wakefield accelerators aimed at achieving TV/m gradients. While Tajima’s original concept \cite{TajimaCavenago1987_XrayAccelerator_PhysRevLett.59.1440} proposed using X-rays injected into a crystalline lattice to generate longitudinal electric wakefields, ultrashort charged particle bunches can also excite electric wakefields. In this case, the energy lost by the driving bunch can be transferred to a properly injected witness bunch, thereby increasing its energy.

Wakefields in graphene can be excited through the collective oscillation of the free electron gas confined over the layer surface (often referred to as plasmons), which is triggered by the driving bunch. Plasmon excitations on the surfaces of graphene layers have been extensively investigated, both experimentally \cite{Kinyanjui2012_EELS_graphene_monolayer,Wachsmuth2013_EELS_monolayer_PhysRevB.88.075433, Eberlein2008_EELS_multilayer_PhysRevB.77.233406} and theoretically \cite{FERMANIANKAMMERER2016_kinetic_model_layers,Allison2009_RPA_graphene_PhysRevB.80.195405,loche2018_molecular_dynamics_layers,radovic2007hydrodynamic_model_layer, RADOVIC2010_hydrodynamic_model_layer_one_fluid,RADOVIC2010_hydrodynamic_model_layer,LI2014_quantum_hydrodynamic_model_layer,BORKA2015_hydrodynamic_model_multilayer,Chaves2017_hydrodynamic_graphene_PhysRevB.96.195438}. On the one hand, these electronic excitations have been studied experimentally by electron energy loss spectroscopy (EELS) in single-layer \cite{Kinyanjui2012_EELS_graphene_monolayer,Wachsmuth2013_EELS_monolayer_PhysRevB.88.075433} and multilayer graphene \cite{Eberlein2008_EELS_multilayer_PhysRevB.77.233406}. On the other hand, plasmon excitations on graphene layers have been theoretically studied employing different approaches, e.g. a kinetic model \cite{FERMANIANKAMMERER2016_kinetic_model_layers}, the random phase approximation \cite{Allison2009_RPA_graphene_PhysRevB.80.195405}, molecular dynamics simulations \cite{loche2018_molecular_dynamics_layers} and hydrodynamic models \cite{ radovic2007hydrodynamic_model_layer, RADOVIC2010_hydrodynamic_model_layer_one_fluid, RADOVIC2010_hydrodynamic_model_layer,LI2014_quantum_hydrodynamic_model_layer, BORKA2015_hydrodynamic_model_multilayer, Chaves2017_hydrodynamic_graphene_PhysRevB.96.195438}. In previous studies, most research focus on magnitudes such as image force, energy loss and stopping power, or at most, they evaluate the induced surface electron density at the graphene layers and/or the induced potential. However, these studies do not consider the excited longitudinal and transverse wakefields, which are of particular interest to us as potential drivers of acceleration and focusing for a given witness charge.

The main innovation of the present work is the derivation of general expressions for the wakefields in a general multilayer graphene (with $N$ layers, where each sheet may have a different surface density and even an insulating substrate may be considered underneath). By using different surface densities, a two-fluid model can be also considered, making distinction between carbon's $\sigma$ and $\pi$ electrons. 
Besides, the possibility that a witness beam could simultaneously experience both acceleration and focusing has been thoroughly analyzed as well.

This article is organized as follows. In Section \ref{Linearized hydrodynamic theory} the linearized hydrodynamic theory is formulated for a multilayer graphene. The general expressions are derived for longitudinal and transverse wakefields excited by the interaction of a point-like charged particle with the graphene layers. In Section \ref{Results and discussion}, the excited wakefields and the stopping power are widely investigated, especially for the case of one and two graphene layers. Finally, the main findings of this work are summarized in Section \ref{Conclusions}. For simplicity, atomic units are used during the derivation of the wakefield expressions, although they are later converted to SI units to present the final results.

\section{Methodology}\label{Linearized hydrodynamic theory}

\begin{figure}[!h]
\includegraphics[width=\columnwidth]{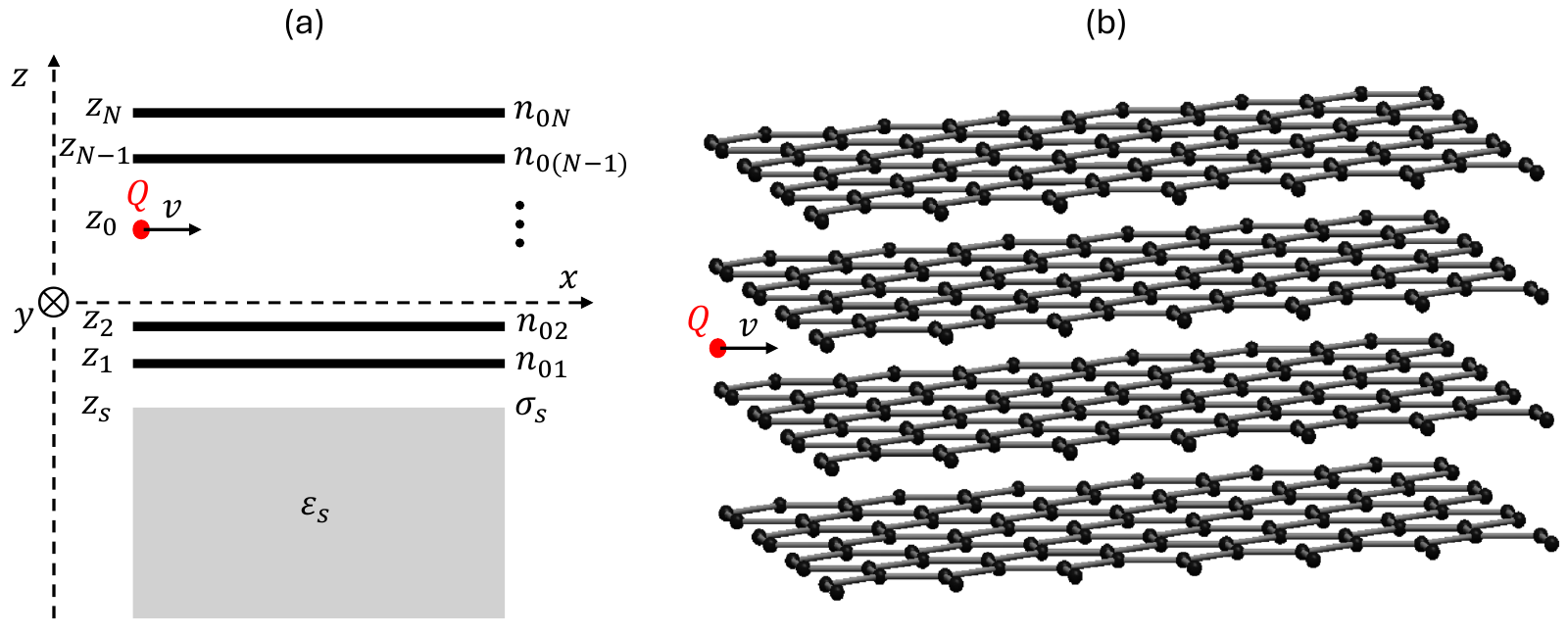}
\centering
\caption{(a) Scheme of the considered charge $Q$ traveling parallel to the $x$-axis in the considered system composed of $N$ graphene layers supported by a substrate with relative permittivity $\varepsilon_s$. (b) Schematic model of the hexagonal lattice of 4 graphene layers.}
\label{fig:Scheme}
\end{figure}

In this work, a linearized hydrodynamic theory is adopted to describe excitations produced by a single charged particle on graphene layers. In this model, we assume $N$ parallel graphene layers located at planes $z_1 \leq z_2 \leq \dotsi \leq z_N$, each treated as infinitesimally thin, and an insulating substrate with relative permittivity $\varepsilon_s$ in the region $z\leq z_s$; the region $z>z_s$ is assumed to be vacuum. The delocalized electrons of the carbon ions are considered as a two-dimensional free-electron gas (Fermi gas) that is confined over the surface of the $j$th layer with a uniform surface density $n_{0j}$. We consider a driving point-like charge $Q$, traveling parallel to the $x$-axis with a constant velocity $v$ (see Figure \ref{fig:Scheme}). Therefore, the driver position as a function of time $t$ is $\mathbf{r}_0(t)=\left(vt, y_0, z_0\right)$. The homogeneous electron gas at each layer, which will be perturbed by the driving charge $Q$, can be modelled as a charged fluid, with velocity fields $\mathbf{u}_j(\mathbf{r}_j ,t)$ and surface density $n(\mathbf{r}_j ,t)=n_{0j}+n_j(\mathbf{r}_j ,t)$, where $\mathbf{r}_j=(x,y,z_j)$ are the coordinates of a point at the surface of the $j$th layer, and $n_j(\mathbf{r}_j ,t)$ is its perturbed density per unit area. In the linearized hydrodynamic model, it is assumed that the perturbed densities $n_j$ and the fluid velocities $\mathbf{u}_j$ are relatively small perturbations. Since the electron gases are confined to each layer, the normal component of the velocity fields $\mathbf{u}_j$ at the surface of the graphene layers is zero. The time scale for ionic motion is significantly slower than that of electronic motion because carbon ions are much heavier than electrons. Thus, for studying wakefield dynamics, ionic motion can be disregarded \cite{Hakimi2018_ionic_motion_PoP_10.1063/1.5016445, Hakimi2020_ionic_motion_nanotube_doi:10.1142/S0217751X19430115}.

In this model, the electronic excitations on the graphene layers are described by the continuity equation,

\begin{equation}\label{eq:continuity_equation}
\frac{\partial n_{j}\left(\mathbf{r}_j, t\right)}{\partial t}+n_{0j} \nabla_{j} \cdot \mathbf{u}_j\left(\mathbf{r}_j, t\right)=0,
\end{equation}

\noindent and the momentum-balance equation of the electron fluid at each surface,

\begin{equation}\label{eq:momentum_balance_equation}
\begin{split}
\frac{\partial \mathbf{u}_j\left(\mathbf{r}_j, t\right)}{\partial t}=\nabla_{j} \Phi\left(\mathbf{r}_j, t\right)-\frac{\alpha_j}{n_{0j}} \nabla_{j} n_j\left(\mathbf{r}_j, t\right)+\frac{\beta}{n_{0j}} \nabla_{j}\left[\nabla_{j}^2 n_j\left(\mathbf{r}_j,t\right)\right]-\gamma_j \mathbf{u}_j\left(\mathbf{r}_j, t\right),
\end{split}
\end{equation}


\noindent where we have retained only the first-order terms in $n_j$ and $\mathbf{u}_j$. In these equations, $\mathbf{r}=(x,y,z)$ is the position vector and $\nabla_j=\hat{\mathbf{x}}\frac{\partial}{\partial x} +\hat{\mathbf{y}}\frac{\partial}{\partial y}$ differentiates only tangentially to the $j$th graphene surface. Equation (\ref{eq:momentum_balance_equation}) shows the sum of four different contributions. The first term on the right-hand side, where $\Phi$ is the electric scalar potential, represents the force exerted on electrons on the $j$th graphene layer, caused by the tangential component of the electric field generated by the driving charge $Q$ and all the consequent perturbed densities. The second term addresses the potential coupling with acoustic modes by defining the parameter $\alpha_j = v^2_F/2$, with $v_F = (2\pi n_{0j})^{1/2}$ representing the Fermi velocity of the two-dimensional electron gas. The third term, where the parameter $\beta = \frac{1}{4}$ has been defined, represents a quantum correction derived from the functional derivative of the Von Weizsacker gradient correction in the equilibrium kinetic energy of the electron fluid \cite{Nejati2009_doi:10.1063/1.3077306} that characterizes single-electron excitations in the electron gas. The final term denotes a frictional force acting on electrons due to their scattering with ionic-lattice charges, where $\gamma_j$ is the damping parameter which corresponds to the $j$th layer. This friction parameter can also be employed as a phenomenological factor to account for the broadening of plasmon resonance in the excitation spectra of various materials \cite{Arista2001IonsPhysRevA.64.032901}.

The equations (\ref{eq:continuity_equation})-(\ref{eq:momentum_balance_equation}) are coupled by the 3D Poisson's equation in free space. The total electric potential is $\Phi=\Phi_{\mathrm{0}}+\Phi_{\mathrm{ind}}$, where $\Phi_0=\frac{Q}{\left\|\mathbf{r}-\mathbf{r}_{\mathbf{0}}\right\|}$ is the Coulomb potential generated by the driving charge and $\Phi_{\mathrm{ind}}$ is the potential created by the perturbation of the electron fluids and the substrate:

\begin{equation}
   \Phi_{\mathrm{ind}}=\sum_{j=1}^N\Phi_{j}+\Phi_{s},
\end{equation}

\noindent where $\Phi_{j}$ and $\Phi_{s}$ are the electric potentials created by the $j$th layer and the substrate, respectively. Thus, the electric potential $\Phi_{j}$ is given by:

\begin{equation}\label{eq:Phi_j}
\Phi_{j}(\mathbf{r}, t)=- \int \mathrm{d}x^{\prime}\mathrm{d}y^{\prime}  \frac{n_j\left(\mathbf{r}_j^{\prime}, t\right)}{\left\|\mathbf{r}-\mathbf{r}_j^{\prime}\right\|},
\end{equation}

\noindent where $\mathbf{r}_j^{\prime}=(x^{\prime},y^{\prime},z_j)$ are the coordinates of a general point at the surface of the $j$th layer. Similarly, the electric potential $\Phi_{s}$ can be calculated by assuming a surface density $\sigma_s$ at $z_s$:

\begin{equation}\label{eq:Phi_s}
\Phi_{s}(\mathbf{r}, t)= \int \mathrm{d}x^{\prime}\mathrm{d}y^{\prime} \frac{\sigma_s\left(\mathbf{r}_s^{\prime}, t\right)}{\left\|\mathbf{r}-\mathbf{r}_s^{\prime}\right\|},
\end{equation}

\noindent with $\mathbf{r}_s^{\prime}=(x^{\prime},y^{\prime},z_s)$. The surface density $\sigma_s$ can be obtained by imposing the continuity of the normal component of the displacement vector at $z_s$, i.e.,

\begin{equation}\label{eq:D_continuity}
\left.\frac{\partial \Phi}{\partial z}\right|_{z=z_s^+}=\left.\varepsilon_{s} \frac{\partial \Phi}{\partial z}\right|_{z=z_s^-}.
\end{equation}

The linearized hydrodynamic model can be solved by  using a Fourier transform with respect to coordinates in the $xy$-plane, $\mathbf{R}=(x,y) \rightarrow \mathbf{k}=(k_x,k_y)$, and time, $t \rightarrow \omega$, which is defined by:

\begin{equation}\label{eq:Fourier}
A(\mathbf{R}, z, t)=\frac{1}{(2 \pi)^3} \int \widetilde{A}(\mathbf{k}, z, \omega) e^{-i(\omega t-\mathbf{k} \cdot \mathbf{R})} d^2 \mathbf{k} d \omega.
\end{equation}

The Fourier transforms of the electric potentials created by the driving charge, the $j$th graphene layer and the substrate are, respectively,

\begin{equation}
\widetilde{\Phi}_{0}(\mathbf{k}, z, \omega)=\frac{(2 \pi)^2 Q \delta(\omega-\mathbf{k} \cdot \mathbf{v})}{k} e^{-k\left|z-z_0\right|},
\end{equation}

\begin{equation}
\widetilde{\Phi}_{j}(\mathbf{k}, z, \omega)=-\frac{2 \pi}{k} \widetilde{n}_j(\mathbf{k}, \omega) e^{-k|z-z_j|},
\end{equation}

\begin{equation}
\widetilde{\Phi}_{s}(\mathbf{k}, z, \omega)=\frac{2 \pi}{k} \widetilde{\sigma}_s(\mathbf{k}, \omega) e^{-k|z-z_s|},
\end{equation}

\noindent where we have defined $k=\sqrt{k_x^2+k_y^2}$.

By imposing Eq. (\ref{eq:D_continuity}), the Fourier transform of the surface density $\widetilde{\sigma}_s$ can be expressed in terms of the Fourier transform of the perturbed densities $\widetilde{n}_j$:

\begin{equation}
\widetilde{\sigma}_s(\mathbf{k}, \omega)=-E\left[2 \pi Q \delta(\omega-\mathbf{k} \cdot \mathbf{v}) e^{-k|z_s-z_0|}-\sum_{j=1}^{N} \widetilde{n}_j(\mathbf{k}, \omega) e^{-k|z_s-z_j|}\right],
\end{equation}

\noindent where we have defined the parameter
$E=\frac{\varepsilon_s-1}{\varepsilon_s+1}$. After eliminating $\mathbf{u}_j$ in (\ref{eq:momentum_balance_equation}) by using the continuity equation and applying the Fourier transform definition, Eq. (\ref{eq:Fourier}), the Fourier transform of the perturbed densities are related by the system of $N$ coupled linear equations

\begin{equation}\label{eq:matrix_n}
S_j(\mathbf{k},\omega)\tilde{n}_j(\mathbf{k},\omega)-\sum_{l=1}^N G_{j l}(\mathbf{k})\tilde{n}_l(\mathbf{k}, \omega)=B_j(\mathbf{k},\omega),
\end{equation}

\noindent where we have defined the following functions:

\begin{equation}\label{eq:S_j}
S_j(\mathbf{k},\omega)=\omega(\omega+i\gamma_j)-\alpha_j k^2-\beta k^4,
\end{equation}

\begin{equation}\label{eq:G_jl}
G_{j l}(\mathbf{k}) = 2\pi n_{0j} k \left[e^{-k|z_j-z_l|}-E e^{-k|z_s-z_j|}e^{-k|z_s-z_l|}\right] ,
\end{equation}

\begin{equation}\label{eq:B_jl}
B_{j}(\mathbf{k},\omega) = -(2\pi)^2 n_{0j} k Q \delta(\omega-\mathbf{k} \cdot \mathbf{v}) \left[e^{-k|z_j-z_0|}-E e^{-k|z_s-z_j|}e^{-k|z_s-z_0|}\right] .
\end{equation}

Once the $\tilde{n}_j$ have been obtained by solving Eq. (\ref{eq:matrix_n}), the perturbed densities and the electric potential can be calculated by applying the definition of the Fourier transform, Eq. (\ref{eq:Fourier}). In particular, using the symmetry properties of the real and imaginary parts of $\tilde{n}_j$ in $\mathbf{k}$, the perturbed densities and the induced potential are given by

\begin{equation}\label{eq:nj}
n_j(\mathbf{R}, z, t)=\frac{4}{(2 \pi)^3} \int_0^\infty \int_0^\infty \text{Re}[\widetilde{n}_j(\mathbf{k}, z, k_x v)e^{ik_x\zeta}]\cos{(k_yy)} dk_xdk_y,
\end{equation}

\begin{equation}\label{eq:phi_ind}
\Phi_\text{ind}(\mathbf{R}, z, t)=\frac{4}{(2 \pi)^3} \int_0^\infty \int_0^\infty \text{Re}[\widetilde{\Phi}_\text{ind}(\mathbf{k}, z, k_x v)e^{ik_x\zeta}]\cos{(k_yy)} dk_xdk_y,
\end{equation}

\noindent where the comoving coordinate $\zeta=x-vt$ has been defined. Finally, the induced wakefields are simply obtained by calculating the corresponding partial derivatives

\begin{equation}
W_x(\mathbf{R}, z, t)=-\frac{\partial \Phi_{\mathrm{ind}}}{\partial x}, \quad W_y(\mathbf{R}, z, t)=-\frac{\partial \Phi_{\mathrm{ind}}}{\partial y}, \quad W_z(\mathbf{R}, z, t)=-\frac{\partial \Phi_{\mathrm{ind}}}{\partial z}.
\end{equation}

\section{Results and discussion} \label{Results and discussion}

In the following calculations, unless otherwise indicated, it is assumed that the point-like charged particle is a proton (i.e. $Q=1$) traveling along the $x$-axis (i.e. $y_0=z_0=0$), the initial uniform surface electron density of each layer can be approximated by $n_{0j}=n_g=1.53\times10^{20}$\,m$^{-2}$ \cite{Ostling1997_surface_density_PhysRevB.55.13980, WangMiskovic2004TwoFluidModelPhysRevB.70.195418} which corresponds to one $\pi$ electron and three $\sigma$ electrons per carbon atom, and the damping parameter is the same in all layers $\gamma\equiv\gamma_j=10^{-3}\,\text{a.u.}\approx 4.13\times10^{13}$\,s$^{-1}$. We have chosen a finite value of the damping parameter $\gamma$ to facilitate the convergence of the integrals, Eqs. (\ref{eq:nj})-(\ref{eq:phi_ind}), but it is small enough so that its effect (i.e. an exponential decay of the wakefields behind the driving particle \cite{Martin-Luna2023_ExcitationWakefieldsSWCNT_NJP}) is practically negligible.

\subsection{Single-layer graphene}\label{sec:One-layer}
\begin{figure}[h!]
{\includegraphics[width=0.5\columnwidth]{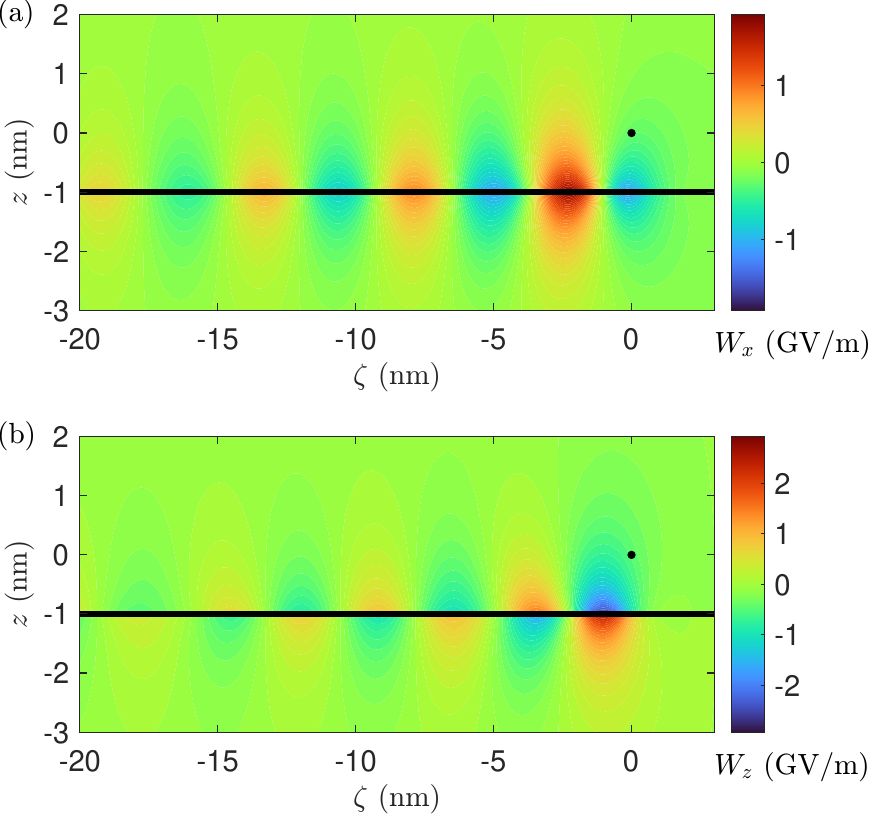}}\hfill
{\includegraphics[width=0.5\columnwidth]{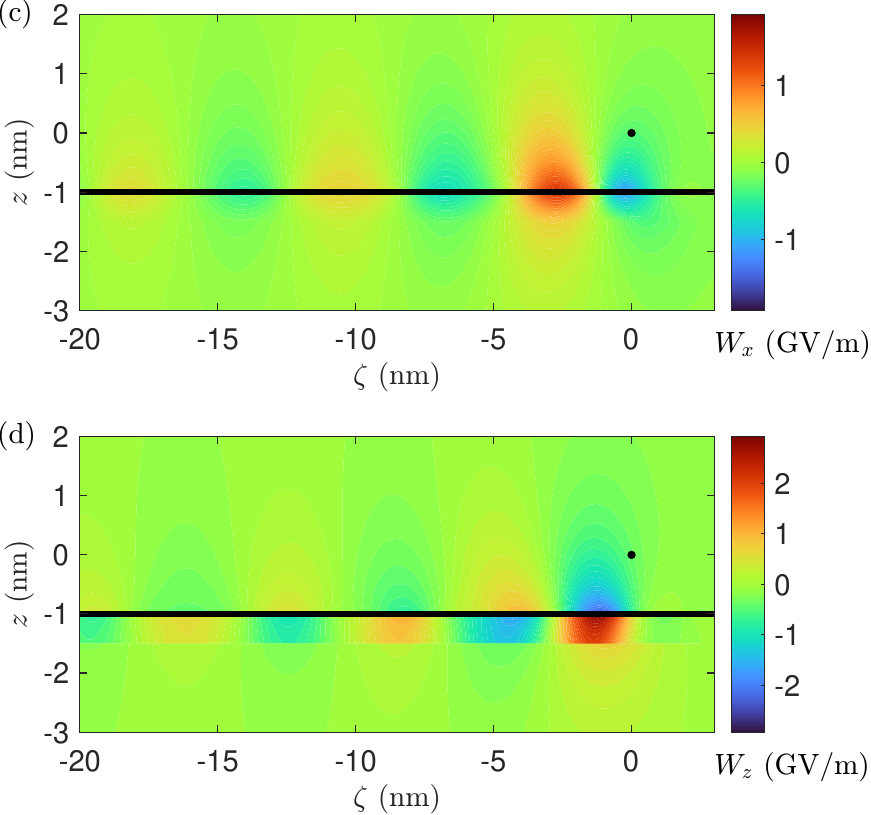}}
\centering
\caption{Induced wakefields (a) $W_x$ and (b) $W_z$ in the $\zeta z$-plane for a proton traveling on the $x$-axis with a velocity $v=0.05c$ above a graphene layer located at $z_1=-1$\,nm. In (a) and (b) we do not consider a substrate; for comparison, in (c) and (d) a SiO$_2$ substrate ($\varepsilon_s=3.9$) is located at $z\leq z_s=-1.5$\,nm. The proton is indicated with a black point and the graphene layer with a black line.}
\label{fig: one_layer_xz_wakefields}
\end{figure}

\begin{figure}[h!]
{\includegraphics[width=0.5\columnwidth]{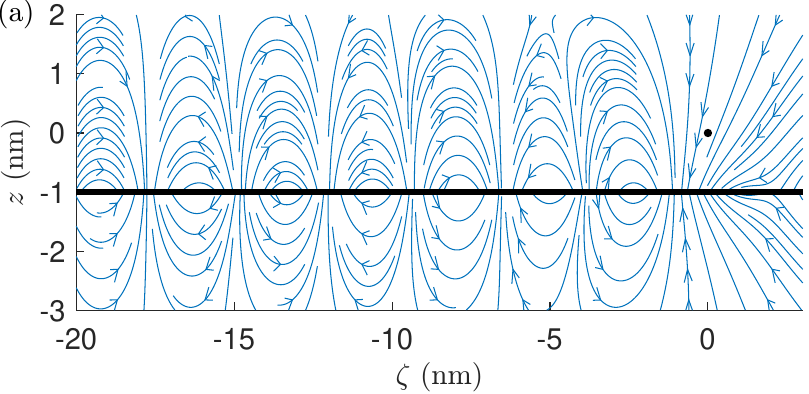}}\hfill
{\includegraphics[width=0.5\columnwidth]{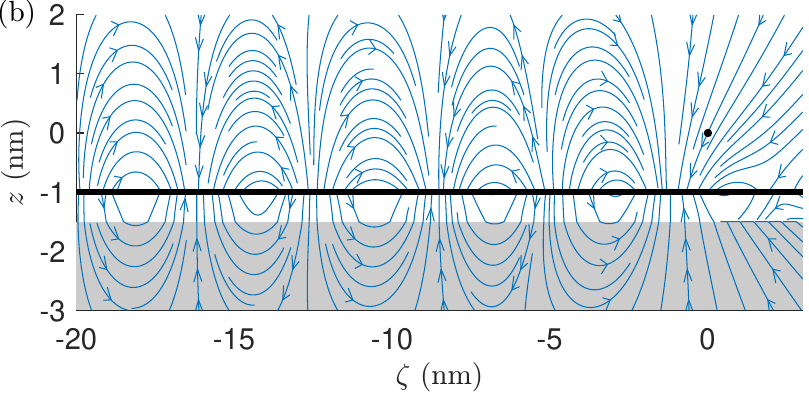}}
\centering
\caption{Electric lines in the $\zeta z$-plane for a proton traveling on the $x$-axis with a velocity $v=0.05c$ above a graphene layer located at $z_1=-1$\,nm. In (a) we do not consider a substrate and in (b) a SiO$_2$ substrate ($\varepsilon_s=3.9$) is located at $z\leq z_s=-1.5$\,nm. The proton is indicated with a black point, the graphene layer with a black line and the substrate with gray background.}
\label{fig: one_layer_xz_streamslice}
\end{figure}

We start by examining the scenario of a single layer. Figure \ref{fig: one_layer_xz_wakefields} illustrates the induced wakefields in the $xz$-plane ($W_y$ is not shown because it is zero), which are excited by a proton traveling at a velocity of $v=0.05c$, positioned 1\,nm from the layer without and with substrate. In both cases, it can be clearly seen the plasmonic excitation in the graphene layer. The longitudinal $W_x$ and transverse $W_z$ wakefields have an offset of $\pi/2$, as can be also seen in Figure \ref{fig: one_layer_xz_streamslice} where the corresponding electric lines are illustrated. If we consider a SiO$_2$ substrate, the wavelength of the plasmonic excitations increases and the wakefields concentrate in the region between the substrate and the graphene layer, producing an enhancement of the transverse wakefield $W_z$, compared to the case without substrate. 

\begin{figure}[h!]
{\includegraphics[width=0.5\columnwidth]{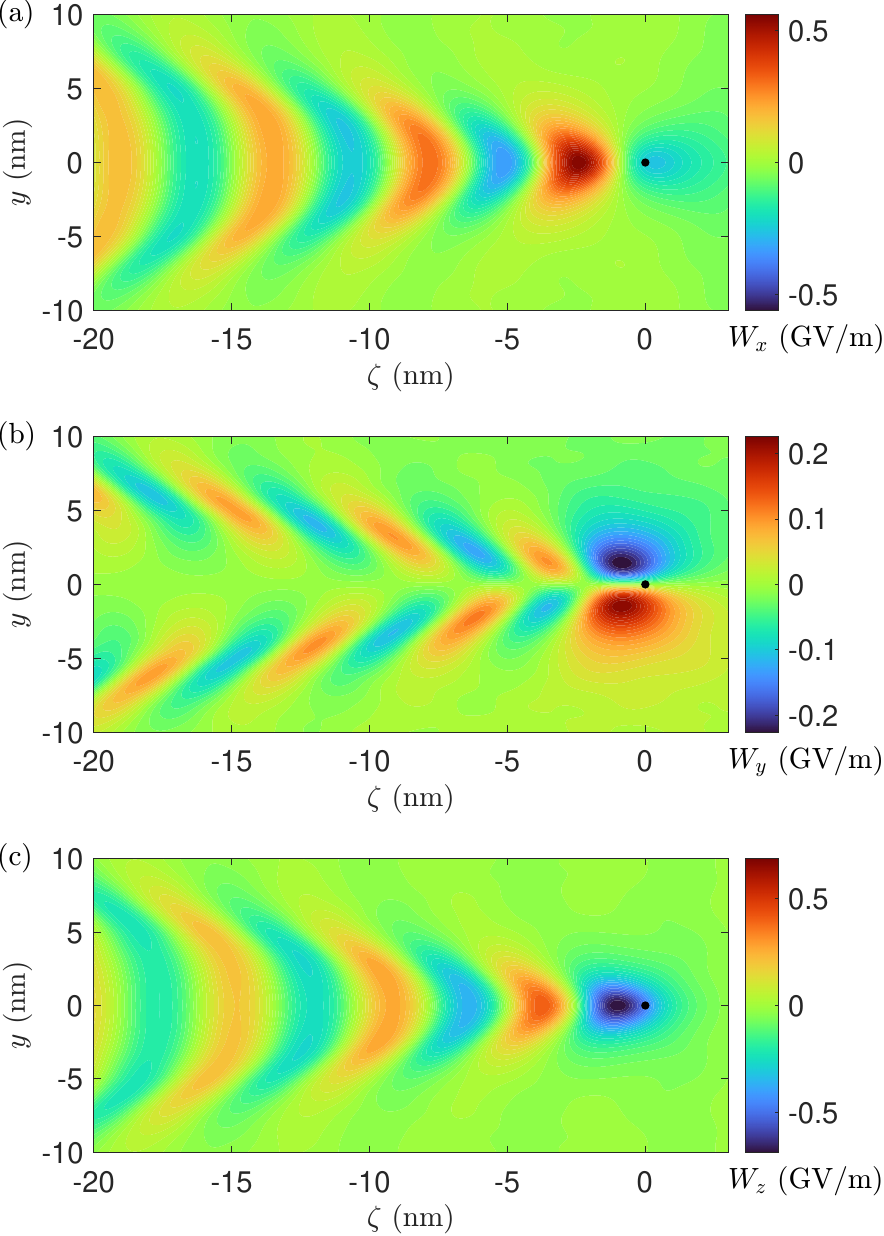}}\hfill
{\includegraphics[width=0.5\columnwidth]{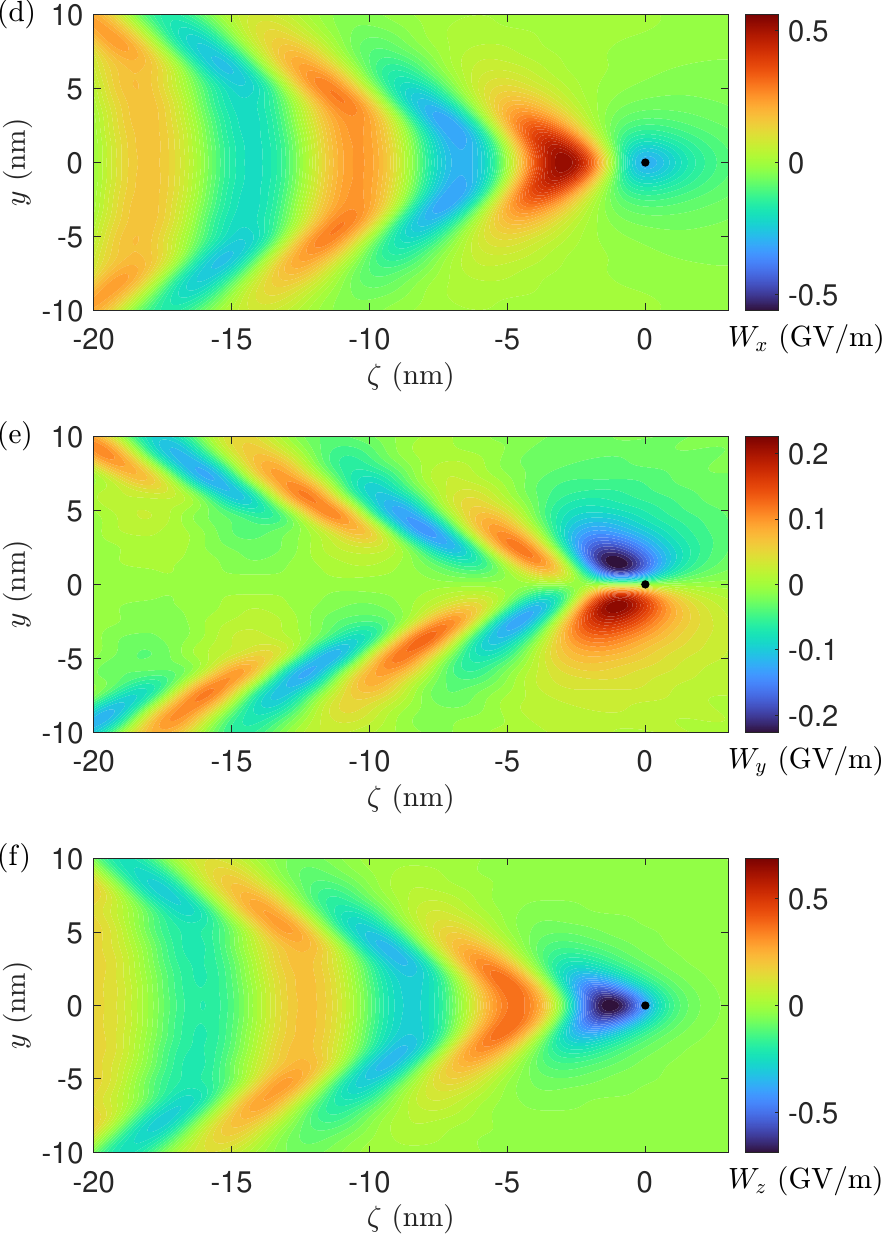}}
\centering
\caption{Induced wakefields (a) $W_x$, (b) $W_y$ and (c) $W_z$ in the $\zeta y$-plane for a proton traveling on the $x$-axis with a velocity $v=0.05c$ above a graphene layer located at $z_1=-1$\,nm. In (a), (b) and (c) we do not consider a substrate; for comparison, in (d), (e) and (f) a SiO$_2$ substrate ($\varepsilon_s=3.9$) is located at $z\leq z_s=-1.5$\,nm. The proton is indicated with a black point.}
\label{fig: one_layer_xy_wakefields}
\end{figure}

\begin{figure}[h!]
{\includegraphics[width=0.5\columnwidth]{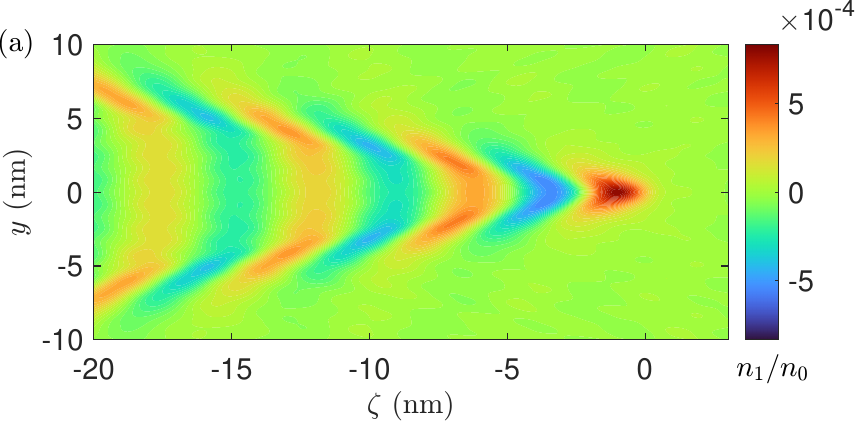}}\hfill
{\includegraphics[width=0.5\columnwidth]{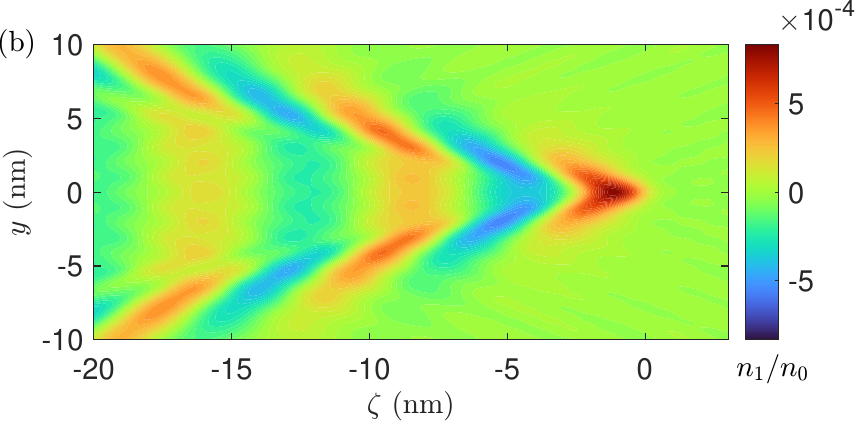}}
\centering
\caption{(a) Perturbed density $n_1/n_0$ in the graphene surface (i.e. at the plane $z_1=-1$\,nm) for a proton traveling on the $x$-axis with a velocity $v=0.05c$. In (a) we do not consider a substrate; for comparison, in (b) a SiO$_2$ substrate ($\varepsilon_s=3.9$) is located at $z\leq z_s=-1.5$\,nm.}
\label{fig: one_layer_xy_n1}
\end{figure}

Figure \ref{fig: one_layer_xy_wakefields} depicts the induced wakefields in the $xy$-plane for the same parameters as in Figure \ref{fig: one_layer_xz_wakefields}. It can be observed that the wakefields spread in the $xy$-plane behind the proton, producing a decay of the intensity of the wakefields with the distance. The induced wakefield $W_y$ has a lower intensity compared to $W_x$ and $W_z$, and exhibits an odd symmetry relative to the $x$-axis. In the case with the SiO$_2$ substrate, the wakefields spread out more spatially in the $xy$-plane. The perturbed density in the graphene layer is depicted in Figure \ref{fig: one_layer_xy_n1}, showing a behavior similar to that of wakefields of Figure \ref{fig: one_layer_xy_wakefields}. It is worth noting that $n_1/n_0 < 10^{-3}\ll 1$, as was assumed in the hydrodynamic model.

\begin{figure}[h!]
{\includegraphics[width=0.495\columnwidth,trim={2.2mm 0.7mm 9.5mm 5mm},clip]{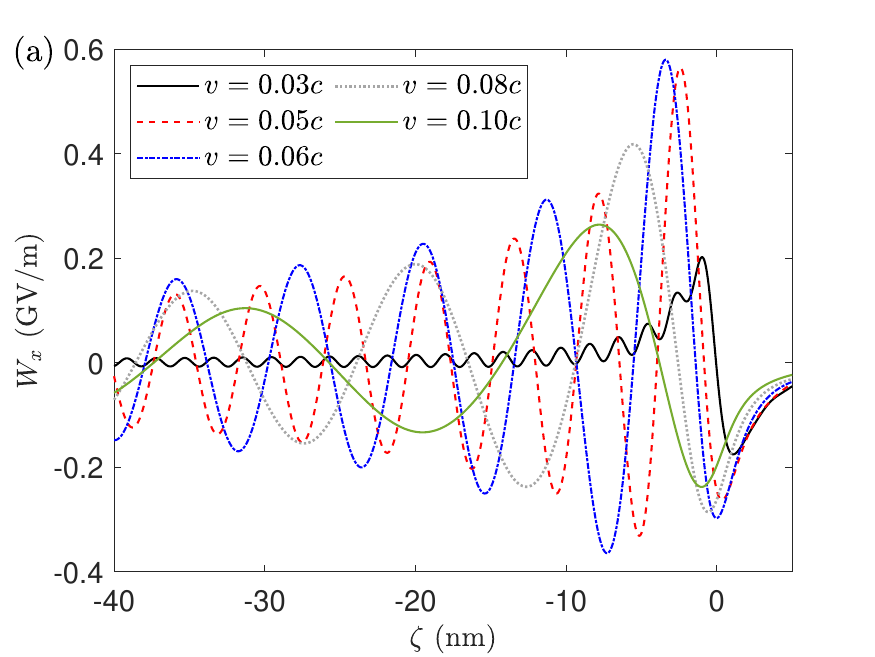}}\hfill
{\includegraphics[width=0.495\columnwidth,trim={2.2mm 0.7mm 9.5mm 5mm},clip]{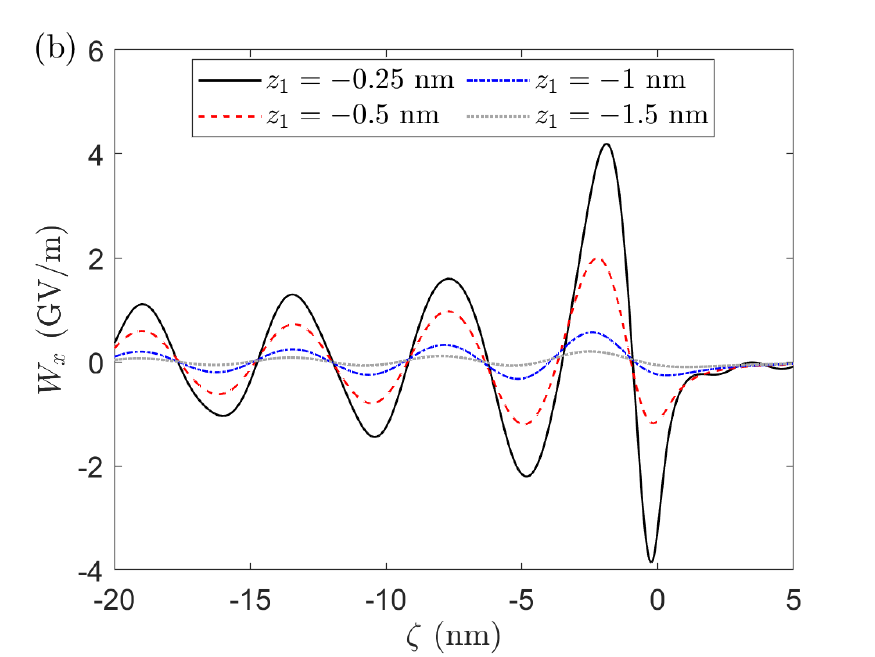}}
\hfill
{\includegraphics[width=0.495\columnwidth,trim={2.2mm 0.7mm 9.5mm 5mm},clip]{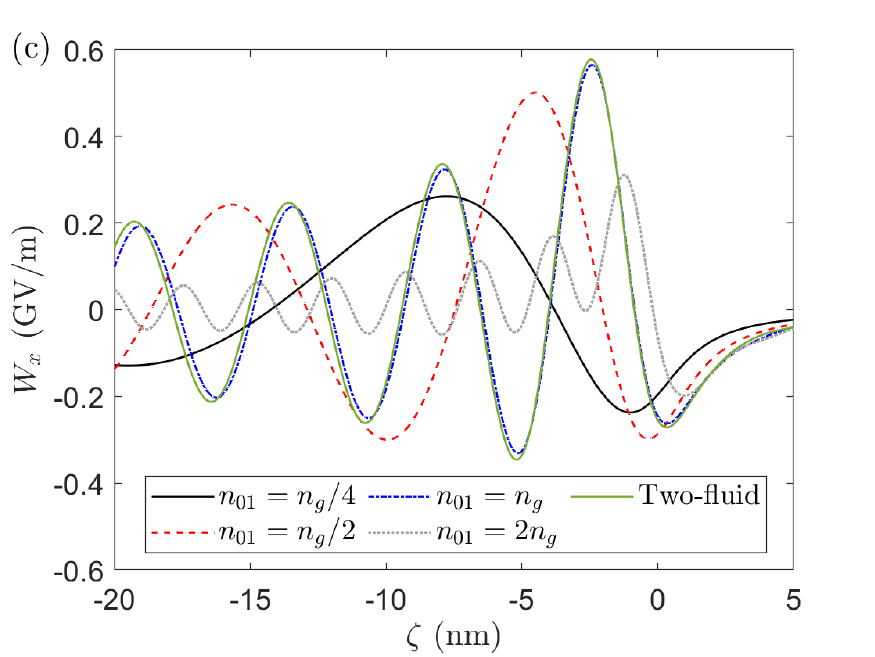}}
\hfill
{\includegraphics[width=0.495\columnwidth,trim={2.2mm 0.7mm 9.5mm 5mm},clip]{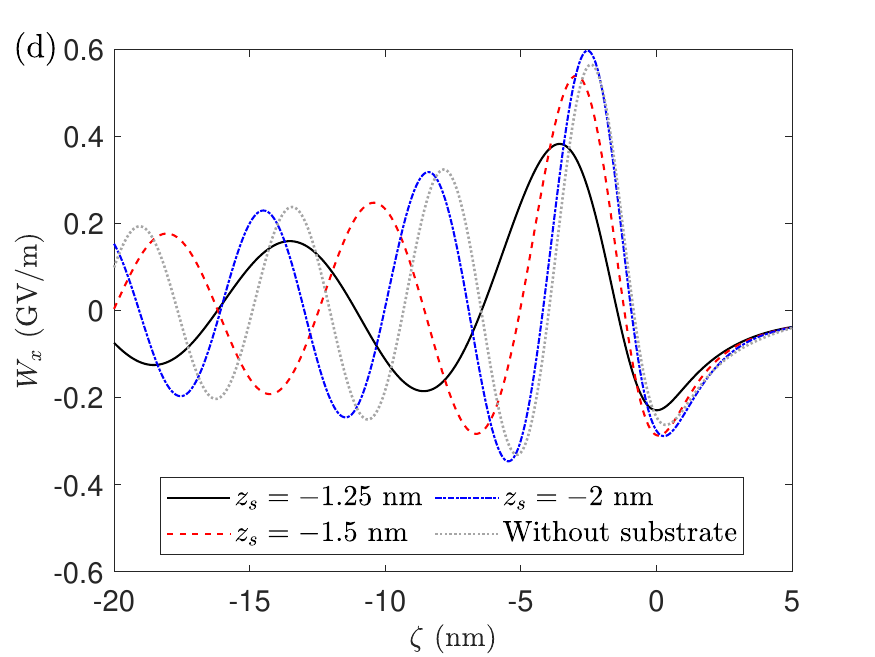}}
\centering
\caption{Induced longitudinal wakefield $W_x$ along the $x$-axis for a proton traveling on the $x$-axis for different (a) velocities $v$, (b) positions of the graphene layer $z_1$, (c) surface densities $n_{01}$ and (d) positions of the SiO$_2$ substrate with $\varepsilon_s=3.9$. In all cases, unless otherwise indicated in the corresponding legend, we do not consider a substrate and use the following parameters: $v=0.05c$, $z_1=-1$\,nm and $n_{01}=n_g$. In (c) it is also considered the two-fluid case where $\sigma$ and $\pi$ electrons are treated as separately fluids, i.e. we assume two layers with $z_1=z_2$ and $n_{01}=0.75n_g$, $n_{02}=0.25n_g$.}
\label{fig: one_layer_wakefields}
\end{figure}

Figure \ref{fig: one_layer_wakefields} shows the induced longitudinal wakefield $W_x$ in the $x$-axis in order to study the effect of each parameter: the driving velocity $v$, the position of the graphene layer $z_1$, the surface density $n_{01}$ and the position of the substrate $z_s$. Figure \ref{fig: one_layer_wakefields}(a) shows that the intensity of the wakefield decreases for low velocities (e.g. $v=0.03c$) and for high velocities 
(e.g. $v=0.10c$). Thus, there is an optimum velocity for which the wakefield intensity is maximum (in this case, $v\approx 0.06c$). Moreover, the wavelength of the plasmonic excitations increases with the velocity $v$. Figure \ref{fig: one_layer_wakefields}(b) illustrates that the plasmonic excitations are more intense the closer the driving particle is to the graphene layer and, besides, the wavelength of the excitations remains approximately constant. Figure \ref{fig: one_layer_wakefields}(c) shows that the wakefields decrease for low and high surface densities, i.e. there is an optimum density (in this case, $n_{01}\approx n_g$). The wavelength of the wakefields decreases with the surface density. For completeness, we have also included the wakefield obtained if we consider a two-fluid model (i.e. treating $\sigma$ and $\pi$ electrons as separated fluids) to show that the results are very similar to the case of the one-fluid model. It is interesting to note that all these behaviors are similar to the case of plasmonic excitations in CNTs \cite{Martin-Luna2023_ExcitationWakefieldsSWCNT_NJP}. Finally, Figure \ref{fig: one_layer_wakefields}(d) depicts the wakefields for different positions of the SiO$_2$ substrate. It can be observed that, in general, the intensity of the wakefield decreases and the wavelength increases the closer the substrate is to the graphene layer.

To conclude the case of a single layer, we are going to study the maximum longitudinal wakefield excited along the $x$-axis ($W_x^{max}$). For completeness, we will also calculate the stopping power (i.e. the energy loss of a channeled particle per unit path length caused by the collective electron excitations), which is given by
\begin{equation}
    \left.S = -Q W_{x} \right|_{\mathbf{r}=\mathbf{r}_0}.
\end{equation}

\begin{figure}[h!]
{\includegraphics[width=0.5\columnwidth,trim={1mm 1.7mm 9.5mm 5
5mm},clip]{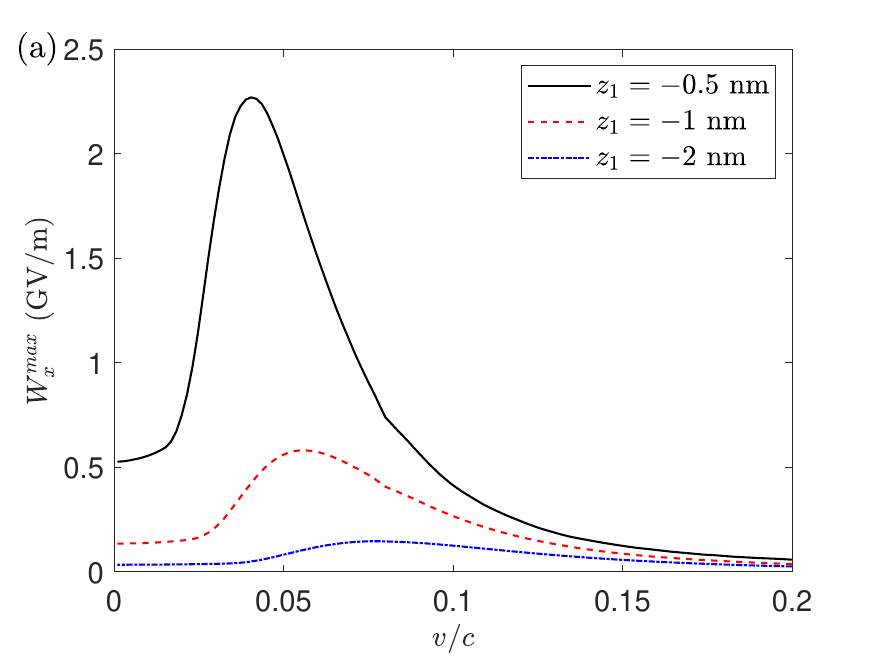}}\hfill
{\includegraphics[width=0.5\columnwidth,trim={1mm 1.7mm 9.5mm 5
5mm},clip]{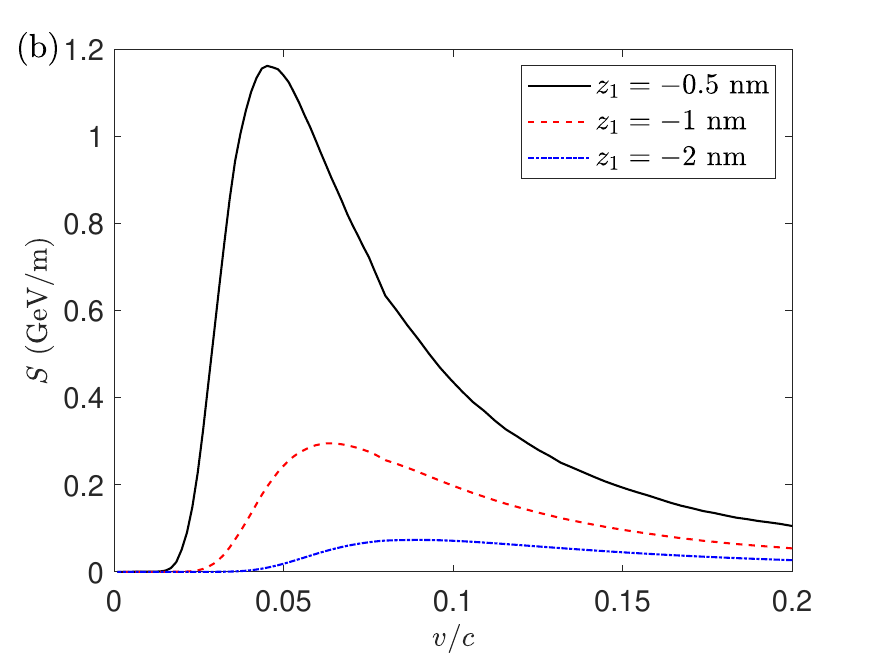}}
\centering
\caption{(a) Maximum longitudinal wakefield $W_x^{max}$ and (b) stopping power as a function of the driving velocity for different values of the position $z_1$ of the graphene layer.  
}
\label{fig: one_layer_max_wake_S_apertures}
\end{figure}

Figure \ref{fig: one_layer_max_wake_S_apertures} shows $W_x^{max}$ and the stopping power as a function of the driving velocity for different values of $z_1$. It can be seen that $W_x^{max}$ decreases when the separation between the driving particle and the graphene layer increases. It is also observed that $W_x^{max}$ and the stopping power have a similar behavior, except at very low velocities where $W_x^{max}$ tends to a finite value and the stopping power to zero. In fact, at these very low velocities, no collective excitation (i.e., an oscillating wakefield behind the driving particle) occurs: the wakefield is only generated near the driving particle and the perturbed density has a bell-shape \cite{RADOVIC2010_hydrodynamic_model_layer, RADOVIC2010_hydrodynamic_model_layer_one_fluid}.

\begin{figure}[h!]
{\includegraphics[width=0.5\columnwidth,trim={1mm 1.7mm 9.5mm 5
5mm},clip]{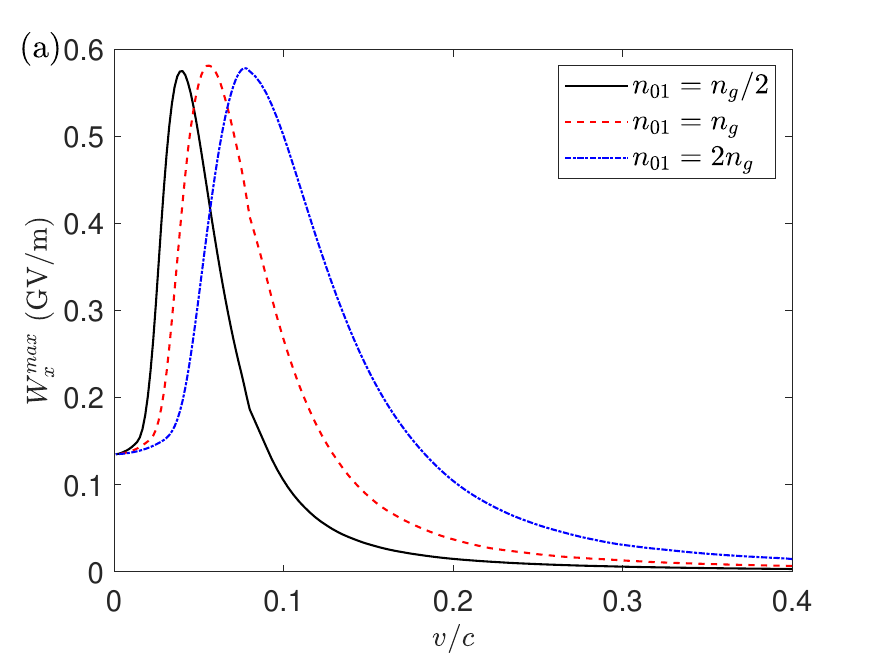}}\hfill
{\includegraphics[width=0.5\columnwidth,trim={1mm 1.7mm 9.5mm 5
5mm},clip]{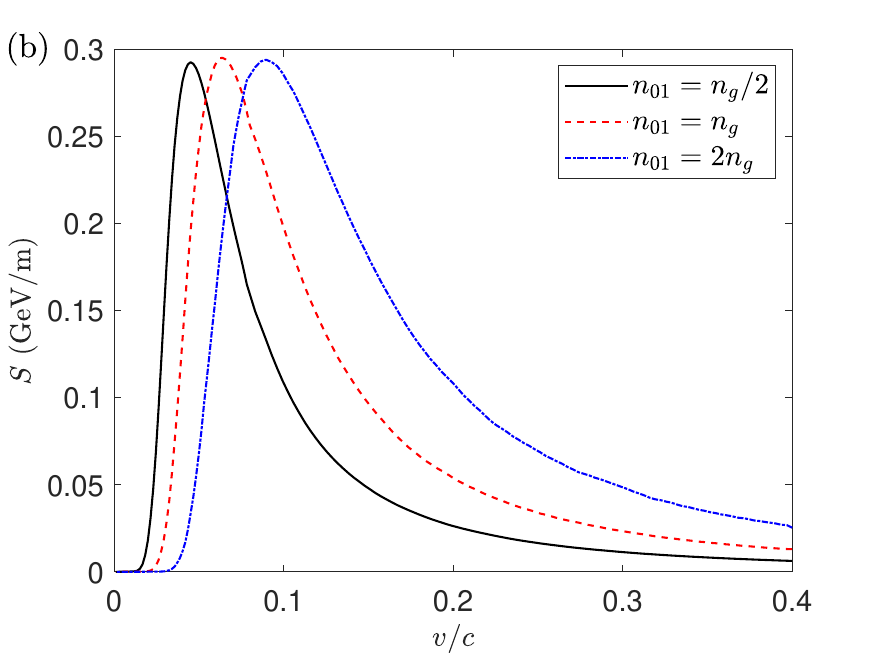}}
\centering
\caption{(a) Maximum longitudinal wakefield $W_x^{max}$ and (b) stopping power as a function of the driving velocity for different values of the surface density $n_{01}$ and $z_1=-1$\,nm.  
}
\label{fig: one_layer_max_wake_S_densities}
\end{figure}

Similarly, Figure \ref{fig: one_layer_max_wake_S_densities} shows $W_x^{max}$ and the stopping power as a function of the driving velocity for different surface densities $n_{01}$. It can be seen that the maximum of $W_x^{max}$ and the stopping power shift to higher velocities as the surface density increases, while their peak values remain approximately constant.

\subsection{Bi-layer graphene}\label{sec:Bi-layer}
In this section we study the case of a proton traveling between two graphene layers with the same surface density $n_0\equiv n_{01}=n_{02}$. 

\begin{figure}[h!]
{\includegraphics[width=0.5\columnwidth]{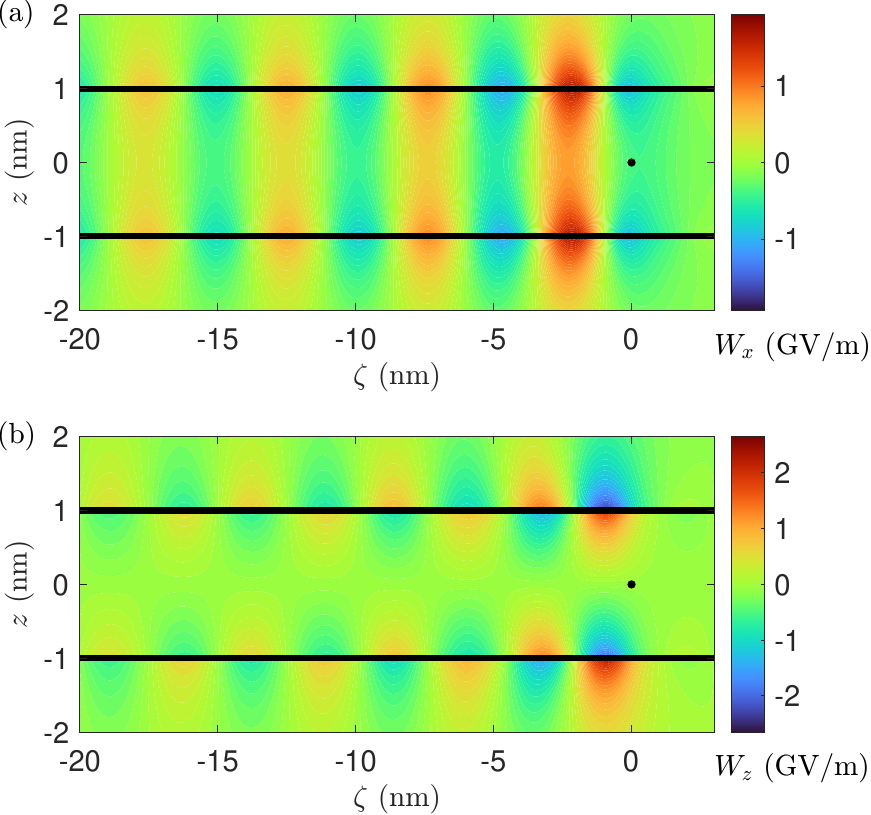}}\hfill
{\includegraphics[width=0.5\columnwidth]{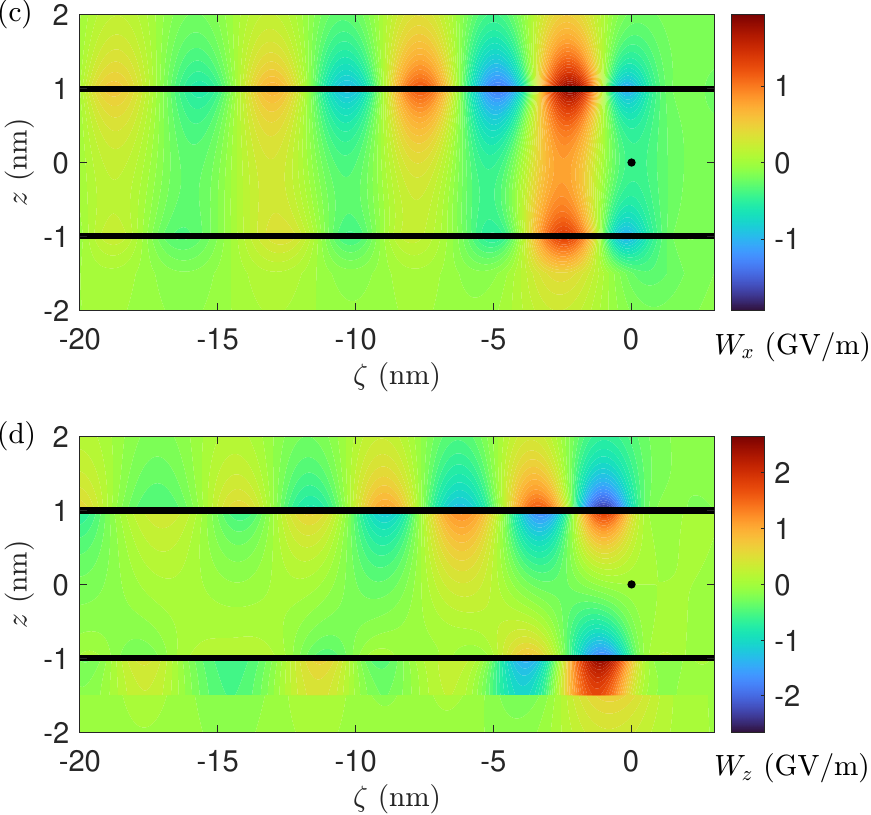}}
\centering
\caption{Induced wakefields (a) $W_x$ and (b) $W_z$ in the $\zeta z$-plane for a proton traveling on the $x$-axis with a velocity $v=0.05c$. The graphene layers are located at $z_1=-1$\,nm and $z_2=1$\,nm. In (a) and (b) we do not consider a substrate; for comparison, in (c) and (d) a SiO$_2$ substrate ($\varepsilon_s=3.9$) is located at $z\leq z_s=-1.5$\,nm. The proton is indicated with a black point and the graphene layers with black lines.}
\label{fig: two_layer_xz_wakefields}
\end{figure}

\begin{figure}[h!]
{\includegraphics[width=0.5\columnwidth]{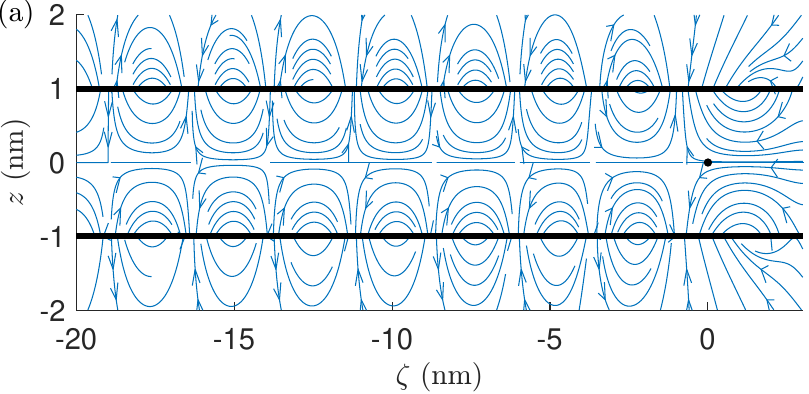}}\hfill
{\includegraphics[width=0.5\columnwidth]{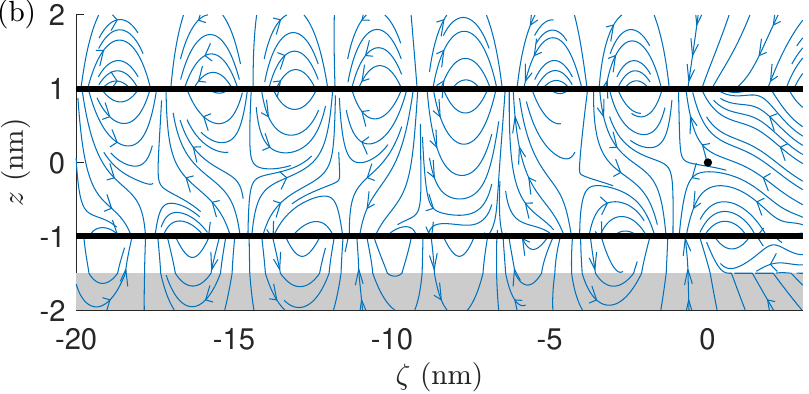}}
\centering
\caption{Electric lines in the $\zeta z$-plane for a proton traveling on the $x$-axis with a velocity $v=0.05c$ between two graphene layers located at $z_1=-1$\,nm and $z_2=1$\,nm. In (a) we do not consider a substrate and in (b) a SiO$_2$ substrate ($\varepsilon_s=3.9$) is located at $z\leq z_s=-1.5$\,nm. The proton is indicated with a black point,  the graphene layers with black lines and the substrate with gray background.}
\label{fig: two_layer_xz_streamslice}
\end{figure}

\begin{figure}[h!]
{\includegraphics[width=0.5\columnwidth]{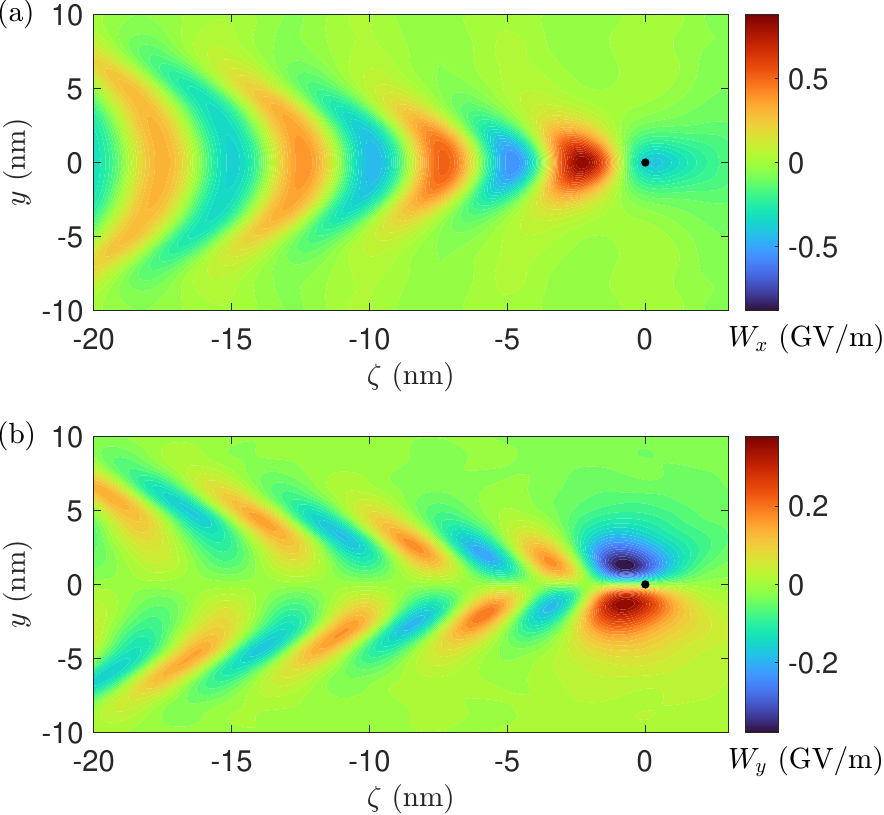}}\hfill
{\includegraphics[width=0.5\columnwidth]{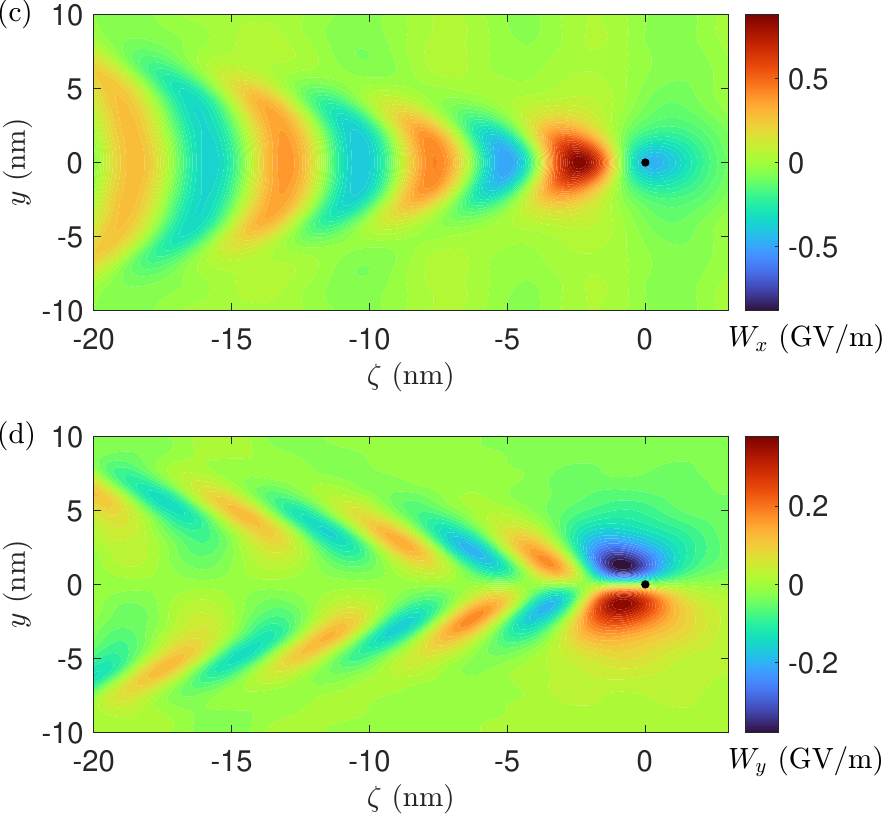}}\hfill
{\includegraphics[width=0.5\columnwidth]{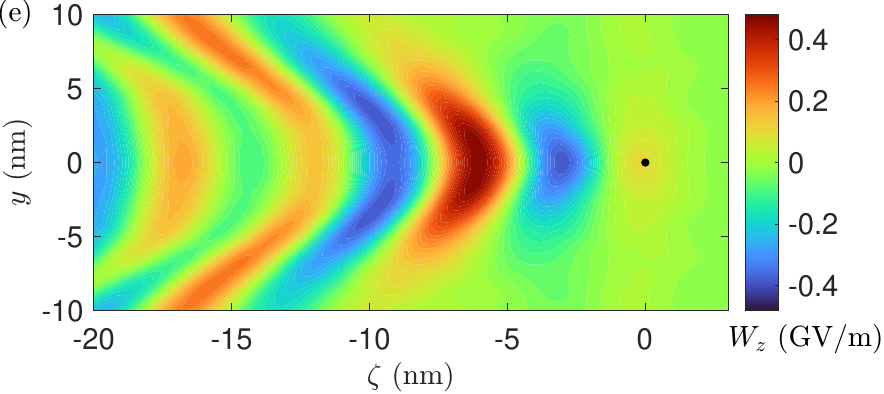}}
\centering
\caption{Induced wakefields (a) $W_x$ and (b) $W_y$ in the $\zeta y$-plane for a proton traveling on the $x$-axis with a velocity $v=0.05c$. The graphene layers are located at $z_1=-1$\,nm and $z_2=1$\,nm. In (a) and (b) we do not consider a substrate; consequently, due to the symmetry, the wakefield $W_z$ is zero. For comparison, we show (c) $W_x$, (d) $W_y$ and (e) $W_z$ considering a SiO$_2$ substrate ($\varepsilon_s=3.9$) located at $z\leq z_s=-1.5$\,nm. The proton is indicated with a black point.}
\label{fig: two_layer_xy_wakefields}
\end{figure}

\begin{figure}[h!]
{\includegraphics[width=0.5\columnwidth,trim={1mm 0.7mm 9.5mm 5mm},clip]{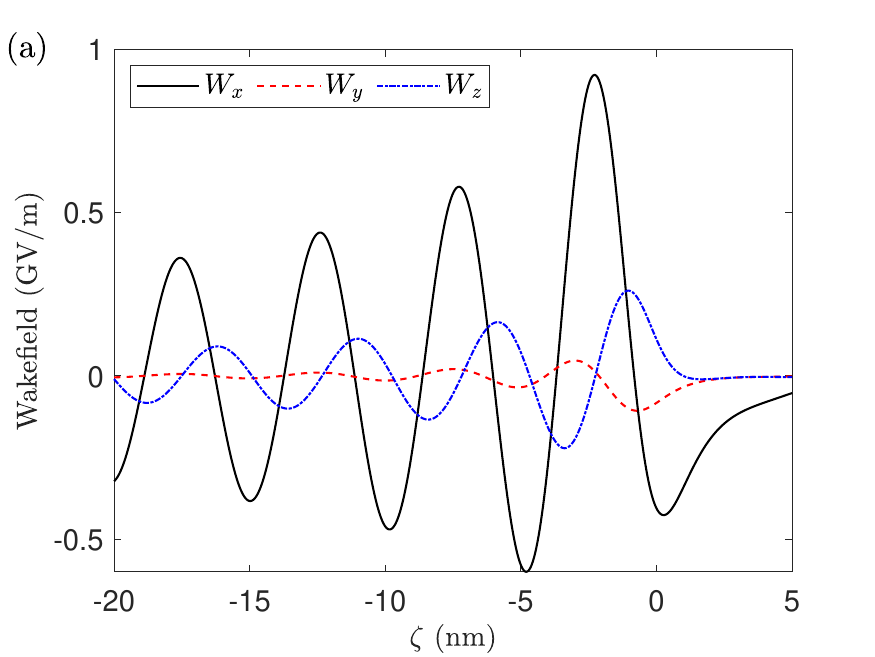}}\hfill
{\includegraphics[width=0.5\columnwidth,trim={1mm 0.7mm 9.5mm 5mm},clip]{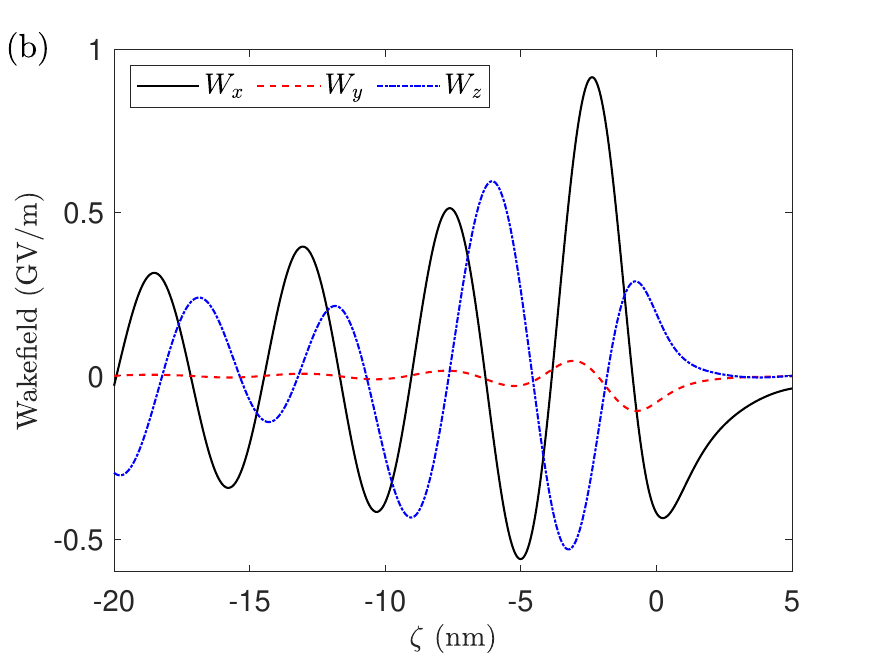}}
\centering
\caption{(a) Induced wakefields along the line $y=z=0.2$\,nm for a proton traveling on the $x$-axis with a velocity $v=0.05c$. The graphene layers are located at $z_1=-1$\,nm and $z_2=1$\,nm. In (b) we consider a SiO$_2$ substrate ($\varepsilon_s=3.9$)  located at $z\leq z_s=-1.5$\,nm.
}
\label{fig: two_layer_acc_foc_defoc}
\end{figure}

\begin{figure}[h!]
{\includegraphics[width=0.5\columnwidth]{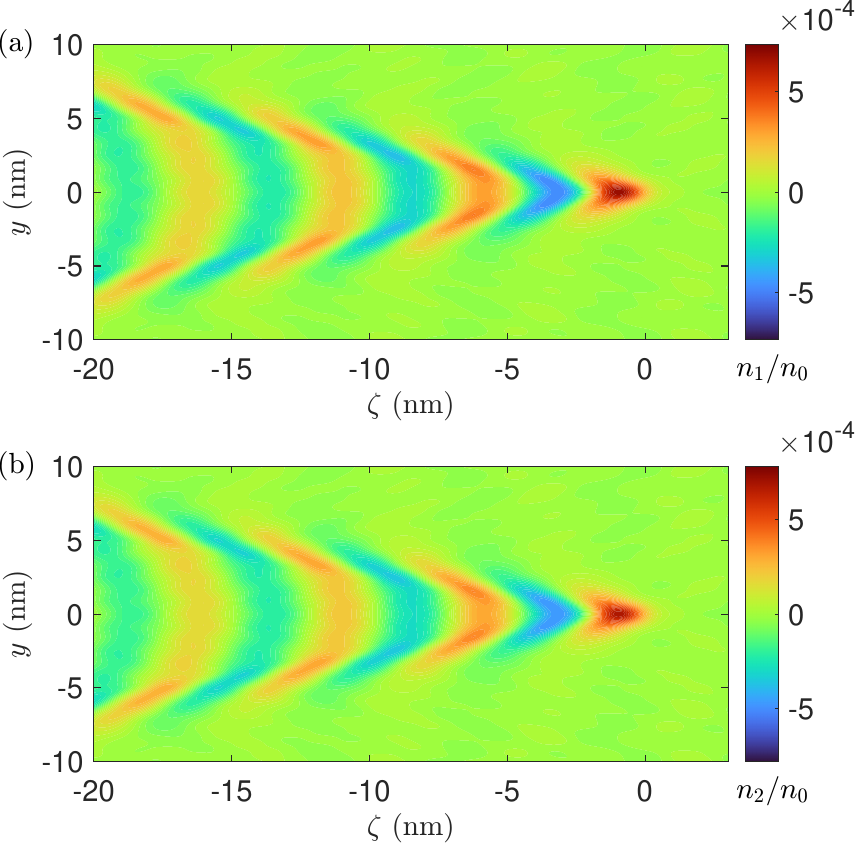}}\hfill
{\includegraphics[width=0.5\columnwidth]{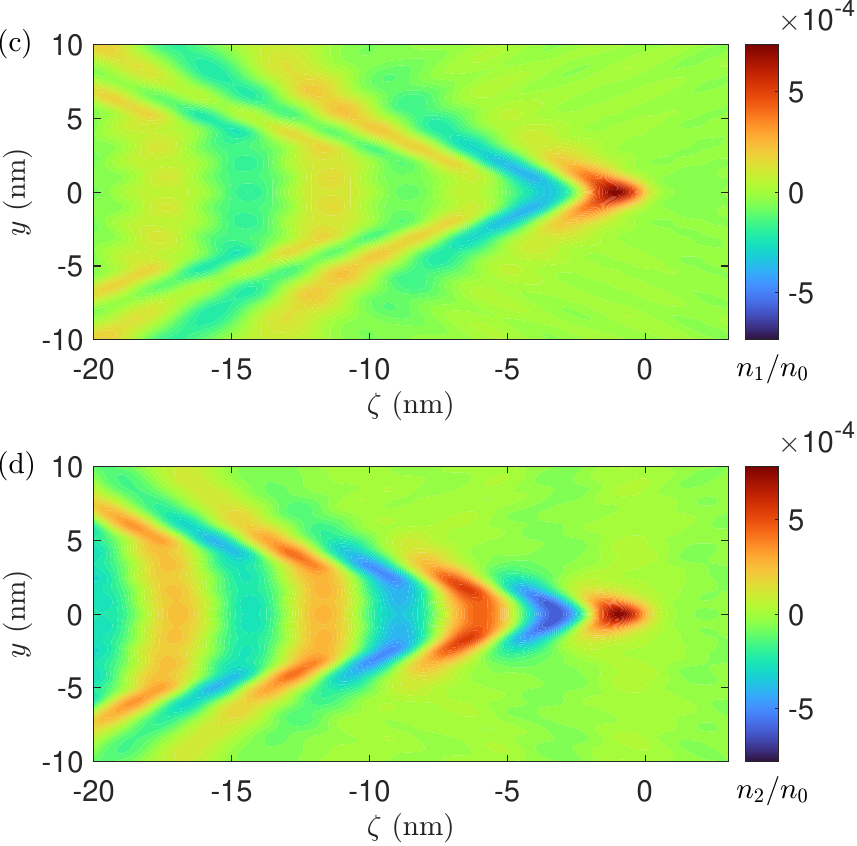}}
\centering
\caption{Perturbed density (a) $n_1/n_0$ and (b) $n_2/n_0$ in the graphene surfaces (i.e. at the planes $z_1=-1$\,nm and $z_2=1$\,nm) for a proton traveling on the $x$-axis with a velocity $v=0.05c$. In (a) and (b) we do not consider a substrate; consequently, due to the symmetry, $n_1=n_2$. For comparison, in (c) and (d) a SiO$_2$ substrate ($\varepsilon_s=3.9$) is located at $z\leq z_s=-1.5$\,nm.}
\label{fig: two_layer_xy_n1}
\end{figure}

Figures \ref{fig: two_layer_xz_wakefields}(a) and (b) show the induced longitudinal $W_x$ and transverse $W_z$ wakefields, respectively, in the $\zeta z$-plane generated by a proton traveling with $v=0.05c$ along the $x$-axis in a bi-layer configuration whose separation is 2\,nm. The wakefield $W_y$ is not shown because is zero in the $\zeta z$-plane. It can be observed that the wakefields $W_x$ and $W_z$ have an offset of $\pi/2$ and their intensities are higher near the graphene layers. Figure \ref{fig: two_layer_xz_streamslice}(a) depicts the corresponding electric lines, where it can be seen that the induced wakefields are similar to the electric fields excited in RF cavities used in conventional particle accelerators. Thus, there are periodical regions where possible witness charged particles can simultaneously experience both acceleration and focusing (if they travel off-axis: $z\neq 0$) in the $\zeta z$-plane. However, in these regions the witness charged particles would be defocused in the $y$-direction (if they travel off-axis: $y\neq 0$), as can be deduced from Figure \ref{fig: two_layer_xy_wakefields}(a) and (b), where the wakefields $W_x$ and $W_y$ are plotted in the $\zeta y$-plane (the wakefield $W_z$ is zero). For a better understanding, Figure \ref{fig: two_layer_acc_foc_defoc} shows the wakefields along the line $y=z=0.2$\,nm, where it can be observed that a witness particle cannot experience both acceleration (i.e. a positive longitudinal wakefield $W_x$) and focusing in the perpendicular plane (i.e. negative transverse wakefields $W_y$ and $W_z$) simultaneously.
In other words, if a beam is focused in one direction it is defocused in the perpendicular direction and vice versa. However, it is worth mentioning that the transverse wakefields are zero along the longitudinal axis, rendering the defocusing of a slightly off-axis witness beam may be practically negligible along short distances. Therefore, this configuration in which a particle travels symmetrically between two graphene layers, represents a potential candidate for particle acceleration purposes. For completeness, Figures \ref{fig: two_layer_xy_n1}(a) and (b) depict the perturbed densities in the graphene surfaces, showing that $n_1=n_2$ and $n_j \ll n_0$, as assumed in the linearized hydrodynamic model.


In order to see how the presence of a substrate affects, Figures \ref{fig: two_layer_xz_wakefields}, \ref{fig: two_layer_xz_streamslice}, \ref{fig: two_layer_xy_wakefields}, \ref{fig: two_layer_acc_foc_defoc} and \ref{fig: two_layer_xy_n1} include in the right column the calculations considering a SiO$_2$ substrate with $\varepsilon_s=3.9$ and $z_s=-1.5$\,nm. Similarly to the single-layer case, there is a concentration of the wakefields in the region between the substrate and the graphene layer below, in particular for $W_z$. Furthermore, the wakefields are no longer symmetrical in the region between the graphene layers, although they are only slightly modified. In particular, the wakefield $W_z$ is not zero (cf. Figure \ref{fig: two_layer_xy_wakefields}(e)) and, besides, the perturbed densities $n_1$ and $n_2$ are different (cf. Figure \ref{fig: two_layer_xy_n1}(c) and (d)) if the substrate is considered. These effects become more important the closer the substrate is to the graphene layers. 

\begin{figure}[h!]
{\includegraphics[width=0.495\columnwidth,trim={2.2mm 0.7mm 9.5mm 5mm},clip]{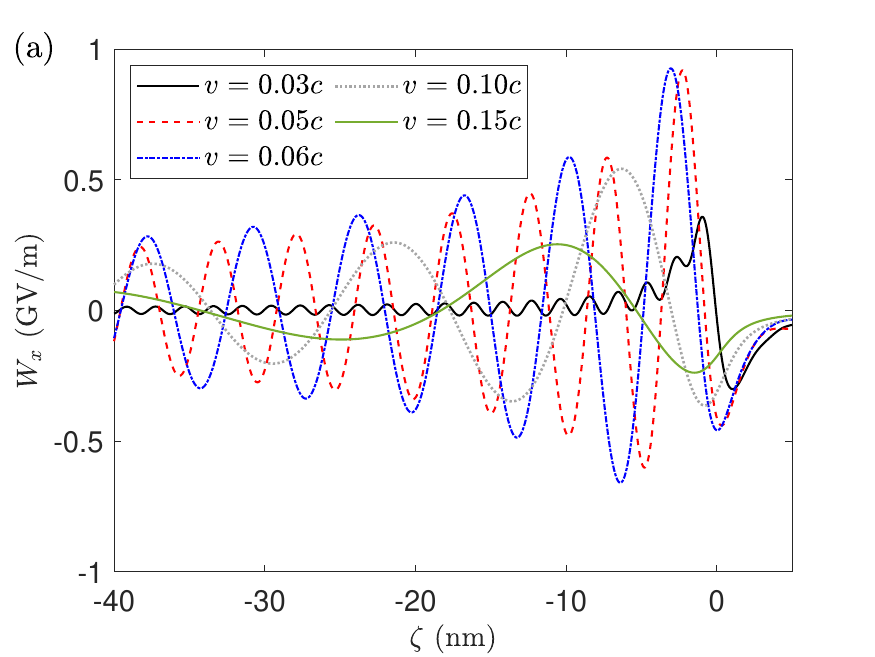}}\hfill
{\includegraphics[width=0.495\columnwidth,trim={2.2mm 0.7mm 9.5mm 5mm},clip]{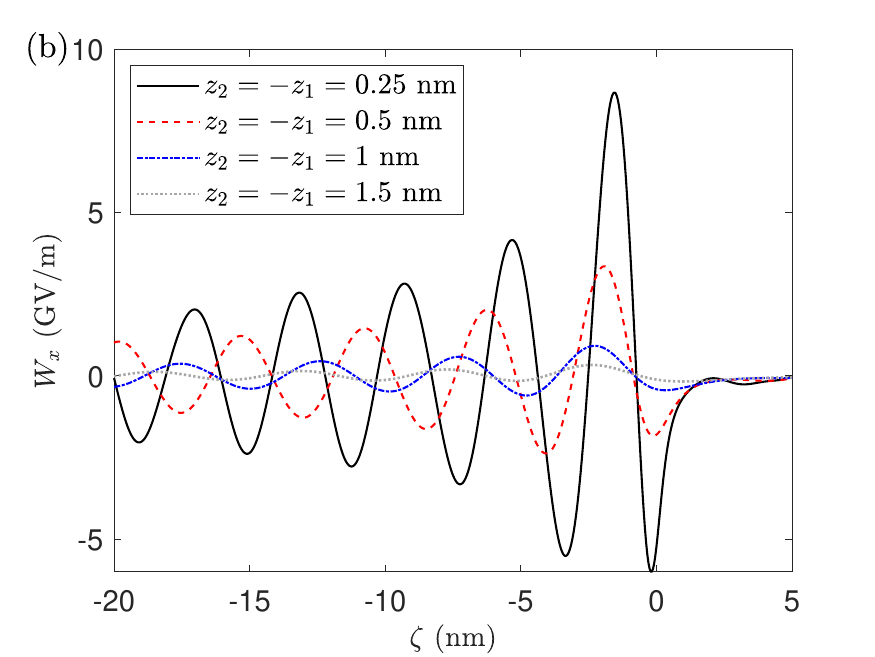}}
\hfill
{\includegraphics[width=0.495\columnwidth,trim={1.8mm 0.7mm 9.5mm 5mm},clip]{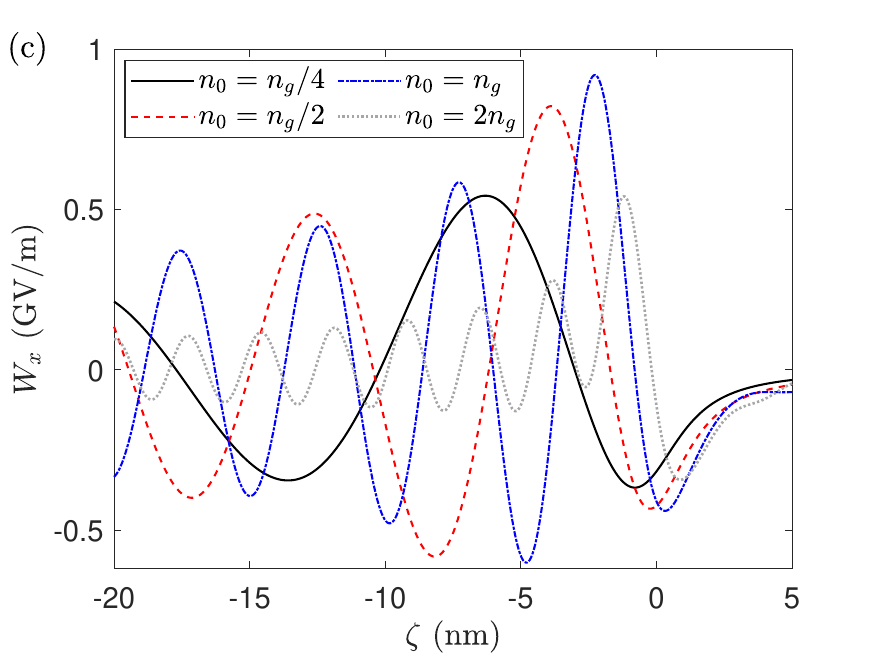}}
\hfill
{\includegraphics[width=0.495\columnwidth,trim={1.8mm 0.7mm 9.5mm 5mm},clip]{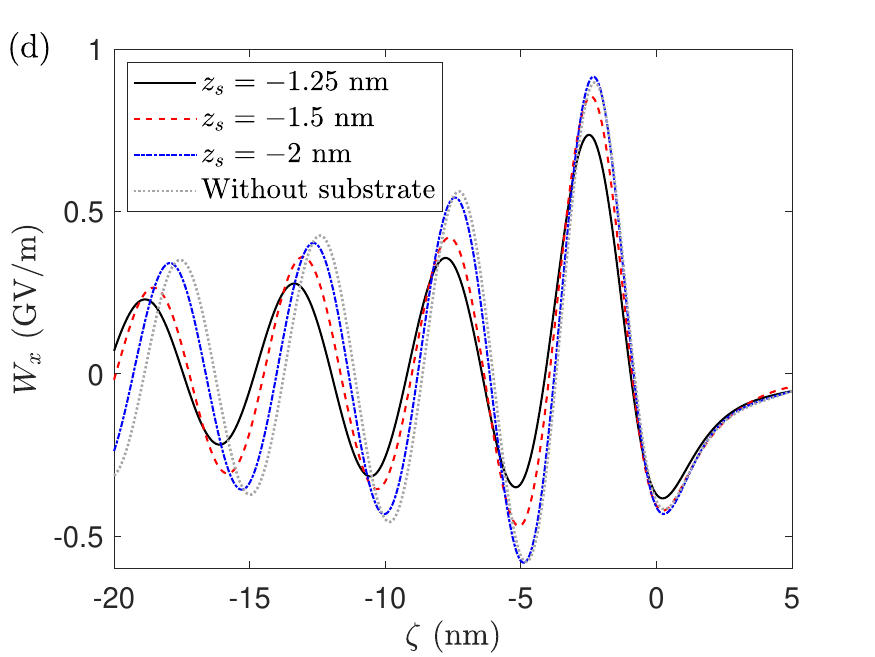}}
\centering
\caption{Induced longitudinal wakefield $W_x$ along the $x$-axis for a proton traveling on the $x$-axis for different (a) velocities $v$, (b) positions of the graphene layers $z_2=-z_1$, (c) surface densities $n_{0}=n_{01}=n_{02}$ and (d) positions of the SiO$_2$ substrate with $\varepsilon_s=3.9$. In all cases, unless otherwise indicated in the corresponding legend, we do not consider a substrate and use the following parameters: $v=0.05c$, $z_1=-1$\,nm and $n_{0}=n_g$. }
\label{fig: two_layer_wakefields}
\end{figure}

As it can be observed in Figure \ref{fig: two_layer_wakefields}, the effect of each parameter (driving velocity, distance between graphene layers, surface density and position of the substrate) is qualitatively similar to the results obtained for the single-layer configuration in Figure \ref{fig: one_layer_wakefields}. The primary distinctions are: (i) the wakefield wavelength increases with greater distance between the graphene layers, and (ii) the substrate has a reduced impact on the wakefields. Furthermore, the intensity of the wakefields is higher than in the single-layer case.

\begin{figure}[h!]
{\includegraphics[width=0.5\columnwidth,trim={1mm 1.7mm 9.5mm 5
5mm},clip]{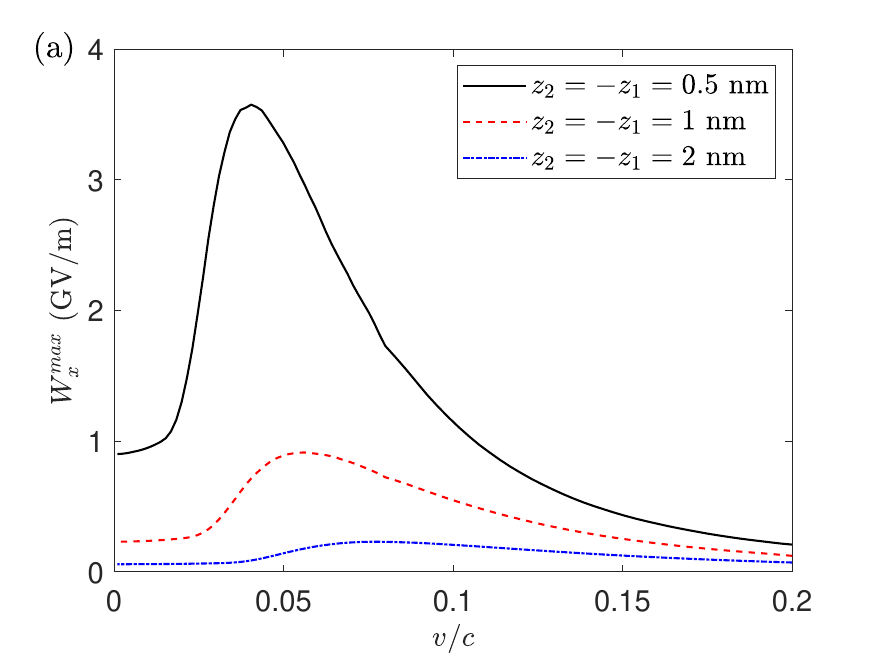}}\hfill
{\includegraphics[width=0.5\columnwidth,trim={1mm 1.7mm 9.5mm 5
5mm},clip]{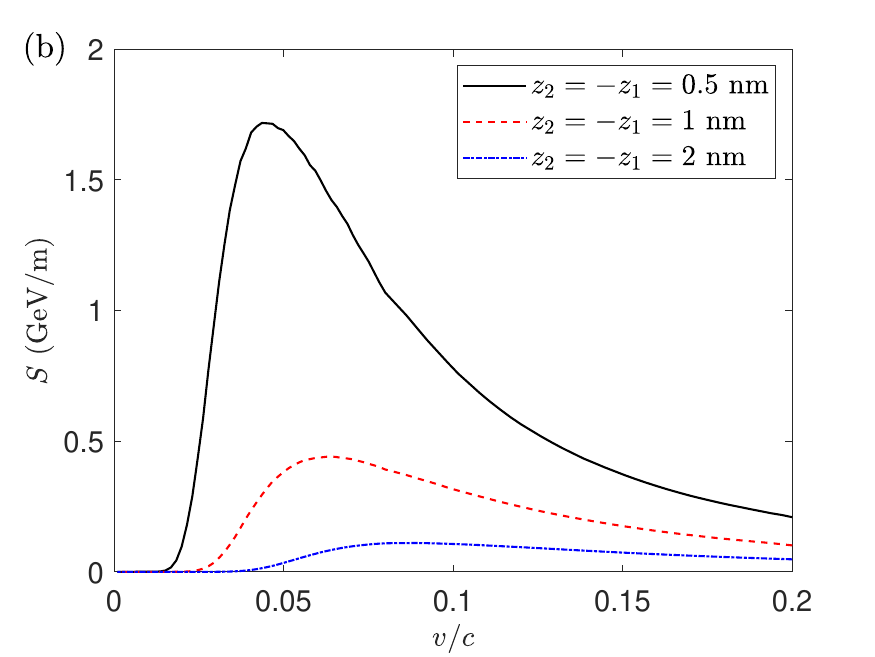}}
\centering
\caption{(a) Maximum longitudinal wakefield $W_x^{max}$ and (b) stopping power as a function of the driving velocity for different values of the position of the graphene layers.  
}
\label{fig: two_layer_max_wake_S_apertures}
\end{figure}

Figure \ref{fig: two_layer_max_wake_S_apertures} shows $W_x^{max}$ and the stopping power as a function of the driving velocity for different values of $z_2=-z_1$. It can be observed that $W_x^{max}$ decreases when the separation between the layer increases. 


\begin{figure}[h!]
{\includegraphics[width=0.5\columnwidth,trim={1mm 1.7mm 9.5mm 5
5mm},clip]{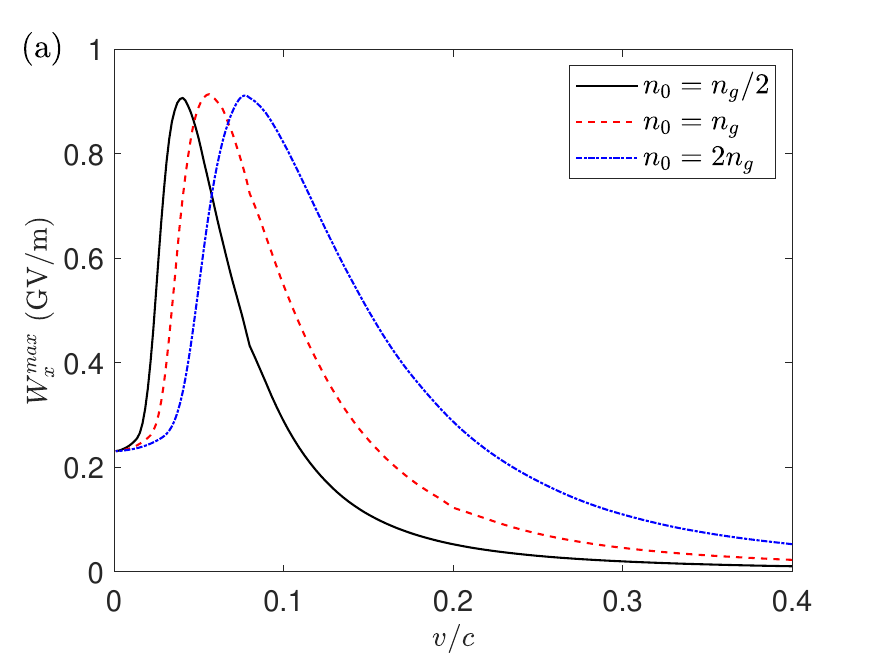}}\hfill
{\includegraphics[width=0.5\columnwidth,trim={1mm 1.7mm 9.5mm 5
5mm},clip]{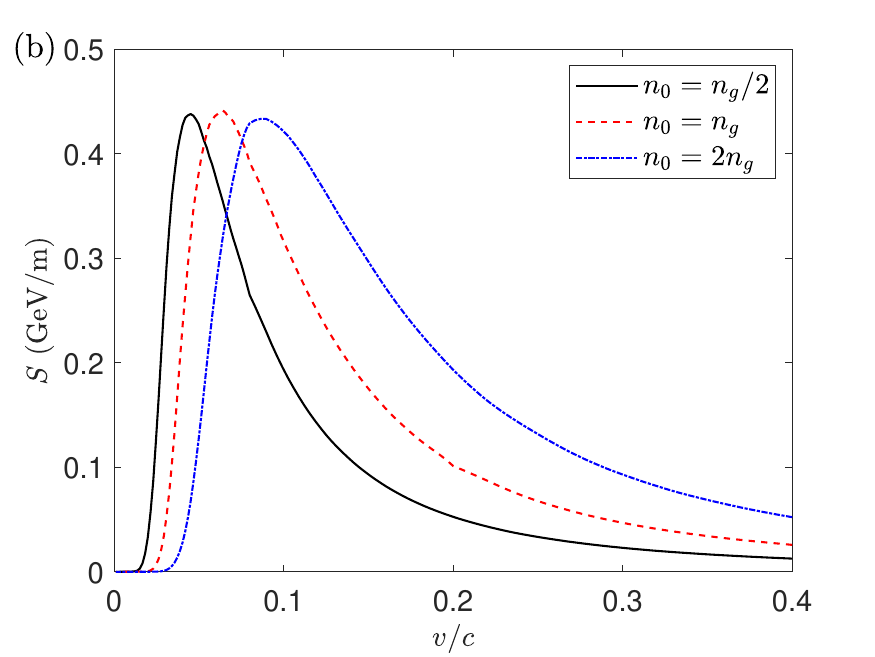}}
\centering
\caption{(a) Maximum longitudinal wakefield $W_x^{max}$ and (b) stopping power as a function of the driving velocity for different values of the surface density $n_{0}$ for $z_2=-z_1=1$\,nm.  
}
\label{fig: two_layer_max_wake_S_densities}
\end{figure}

Similarly, Figure \ref{fig: two_layer_max_wake_S_densities} depicts $W_x^{max}$ and the stopping power as a function of the driving velocity for different surface densities $n_{0}$. As in the case with a single layer, the maximum of $W_x^{max}$ and the stopping power shift to higher velocities as $n_0$ increases, whereas their peak values remain approximately constant.

\subsection{Multilayer graphene}\label{sec: multi-layer}

\begin{figure}[h!]
{\includegraphics[width=0.5\columnwidth]{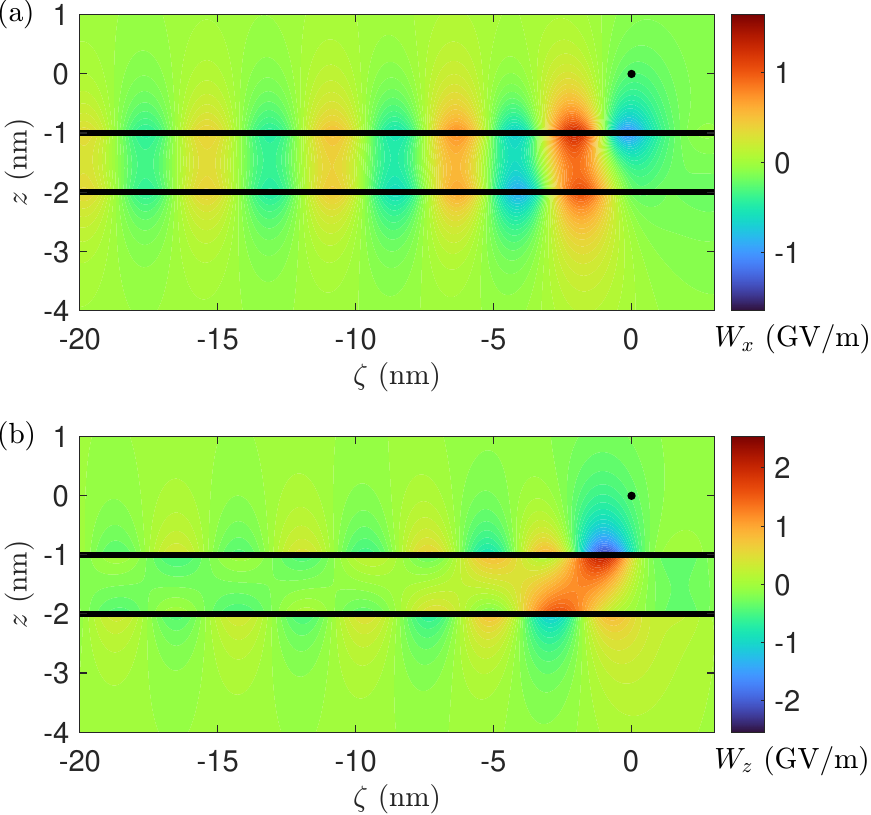}}\hfill
{\includegraphics[width=0.5\columnwidth]{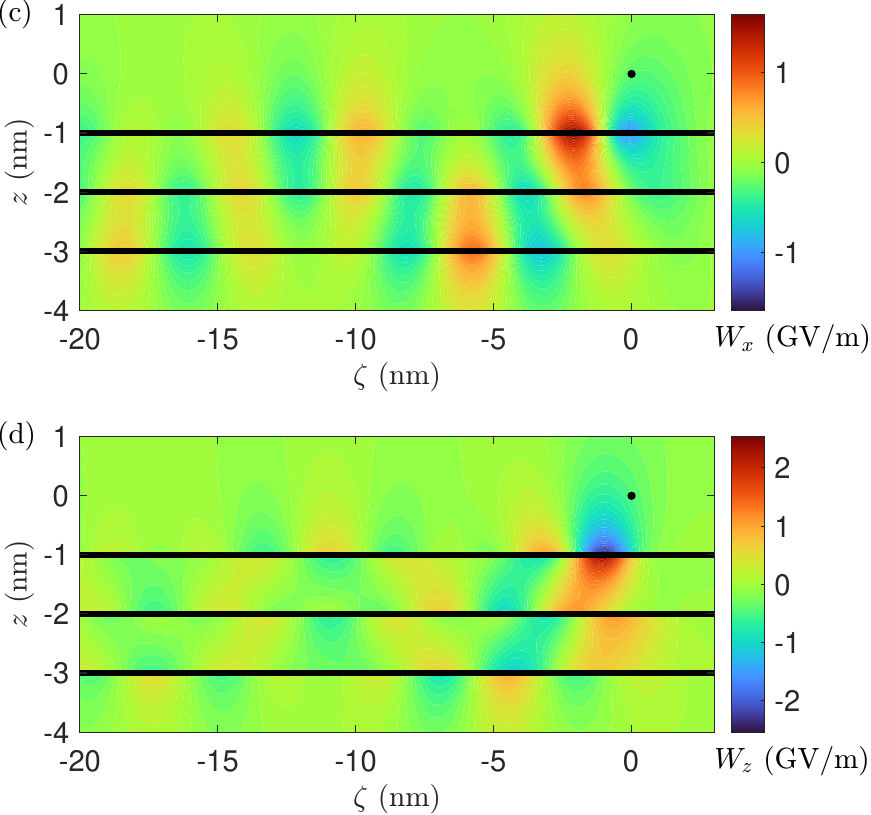}}
\centering
\caption{Induced wakefields (a) $W_x$ and (b) $W_z$ in the $\zeta z$-plane for a proton traveling on the $x$-axis with a velocity $v=0.05c$ above two graphene layers located at $z_1=-2$\,nm and $z_2=-1$\,nm. For comparison, in (c) and (d) we consider three layers at $z_1=-3$\,nm, $z_2=-2$\,nm, and $z_3=-1$\,nm. The proton is indicated with a black point and the graphene layers with black lines.}
\label{fig: layers_abajo_xz_wakefields}
\end{figure}

We start this section by analyzing the case of a proton traveling above $N$ graphene layers located at $z_j=(j-1-N)d$ ($j=1,...,N$), where $d$ is the inter-layer distance. On the one hand, Figure \ref{fig: layers_abajo_xz_wakefields} depicts the induced wakefields in the $\zeta x$-plane for two and three layers, respectively, with $d=1$\,nm. In the case of two layers, the wakefields between the layers resemble the bi-layer configuration described in Section 
\ref{sec:Bi-layer}, although the wakefields are not symmetrical, especially the transverse wakefield $W_z$. With three layers, the wakefields exhibit a more complex pattern, particularly in the region between the layers.

\begin{figure}[h!]
{\includegraphics[width=0.5\columnwidth,trim={1mm 1.7mm 9.5mm 5
5mm},clip]{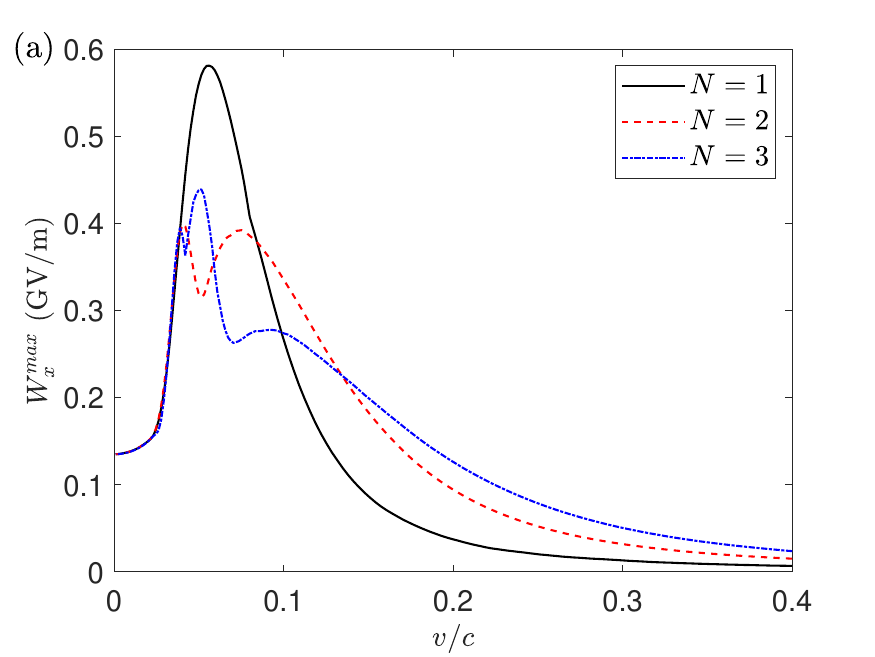}}\hfill
{\includegraphics[width=0.5\columnwidth,trim={0.7mm 1.7mm 9.5mm 5
5mm},clip]{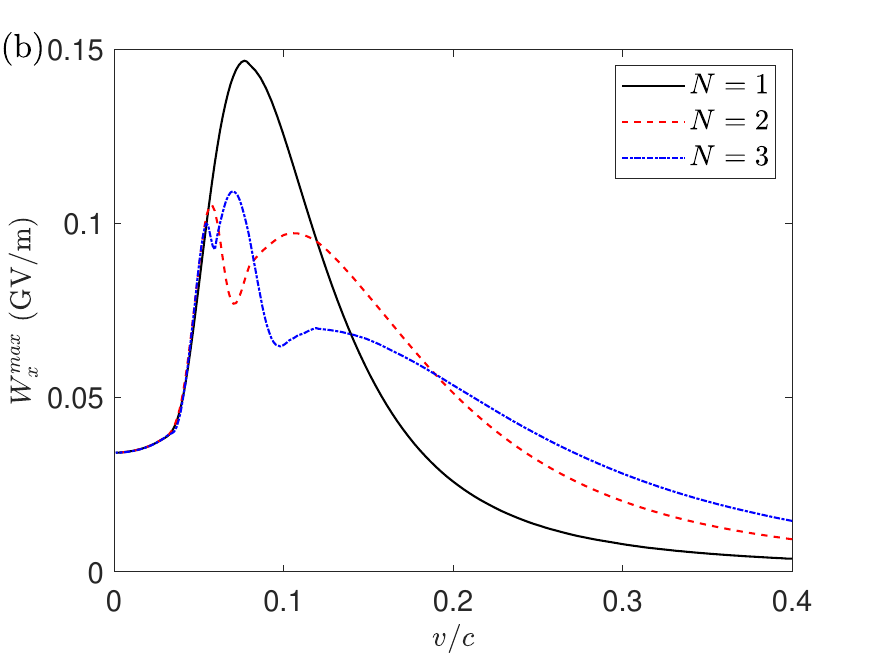}}
\centering
\caption{Maximum longitudinal wakefield $W_x^{max}$ as a function of the driving velocity for different number of graphene layers $N$. The graphene layers are located at $z_j=(j-1-N)d$, where $d$ is the inter-layer distance: (a) $d=1$\,nm and (b) $d=2$\,nm. 
}
\label{fig: N_layer_max_wake}
\end{figure}

On the other hand, Figure \ref{fig: N_layer_max_wake} compares the maximum wakefield $W_x^{max}$ as a function of the velocity for a different number of graphene layers. It can be observed that as $N$ increases, more peaks appear, becoming increasingly narrower. This behavior is similar to that observed in multi-walled CNTs \cite{CHUNG2007_MWCNT_theory_RadPhysChem}. The maximum of $W_x^{max}$ is obtained for a single layer, but as $N$ increases, $W_x^{max}$ becomes greater at higher velocities. When the interplanar distance $d$ increases, $W_x^{max}$ decreases since the distance between the proton and the nearest graphene layer increases.

\begin{figure}[h!]
{\includegraphics[width=0.5\columnwidth]{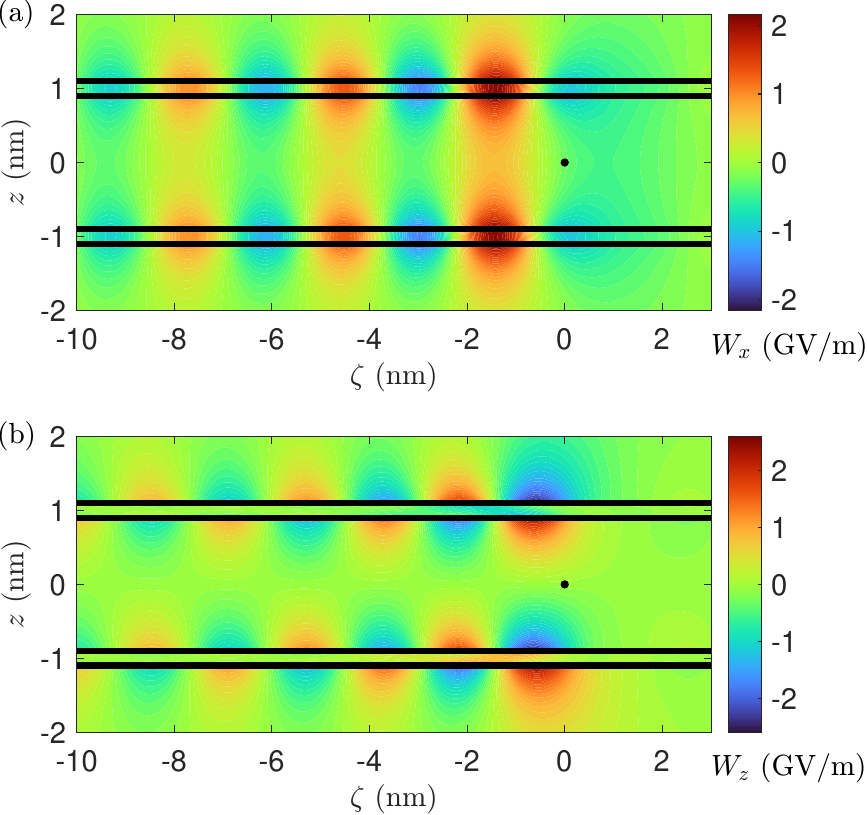}}\hfill
{\includegraphics[width=0.5\columnwidth]{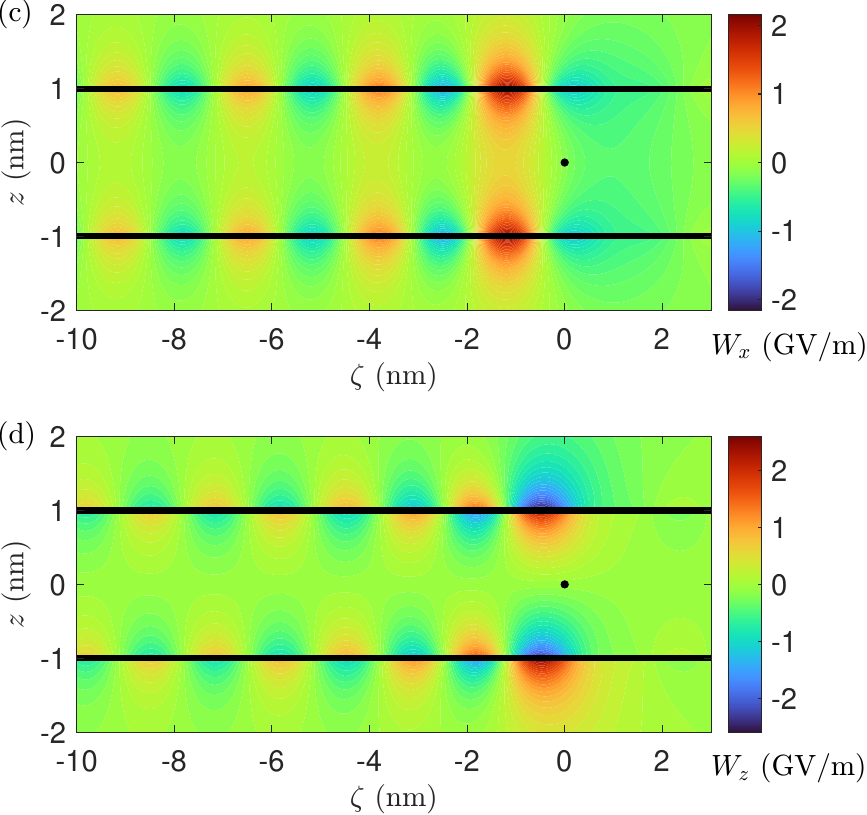}}
\centering
\caption{Induced wakefields (a) $W_x$ and (b) $W_z$ in the $\zeta z$-plane for a proton traveling on the $x$-axis with a velocity $v=0.05c$. The graphene layers are located at $z_4=-z_1=1.1$\,nm and $z_3=-z_2=0.9$\,nm. For comparison, in (c) and (d) we consider two layers at $z_2=-z_1=1$\,nm with $n_0=2n_g$. The proton is indicated with a black point and the graphene layers with black lines.}
\label{fig: 4_layer_xz_wakefields}
\end{figure}

\begin{figure}[h!]
{\includegraphics[width=0.5\columnwidth,trim={1mm 1.7mm 9.5mm 5
5mm},clip]{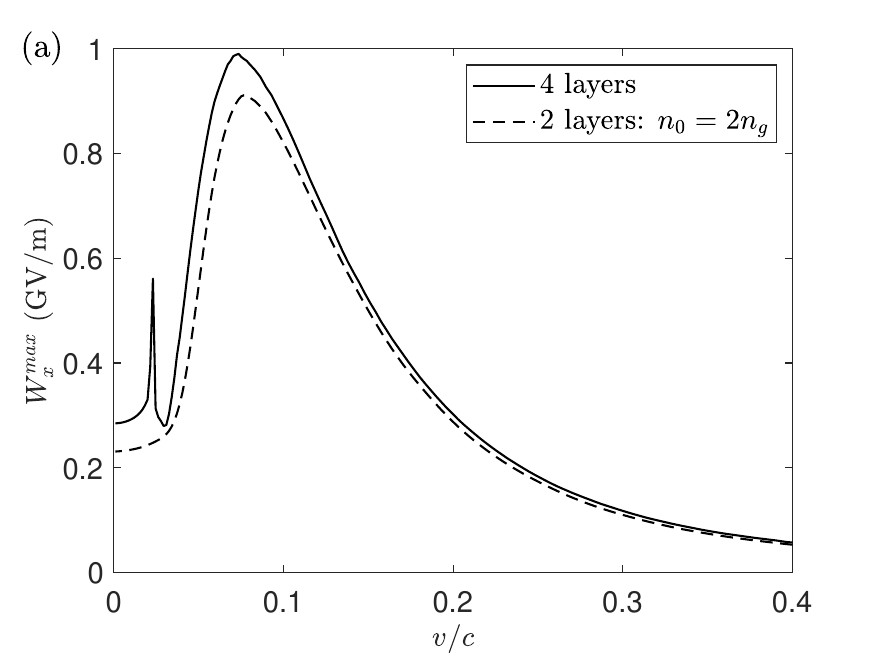}}\hfill
{\includegraphics[width=0.5\columnwidth,trim={1mm 1.7mm 9.5mm 5
5mm},clip]{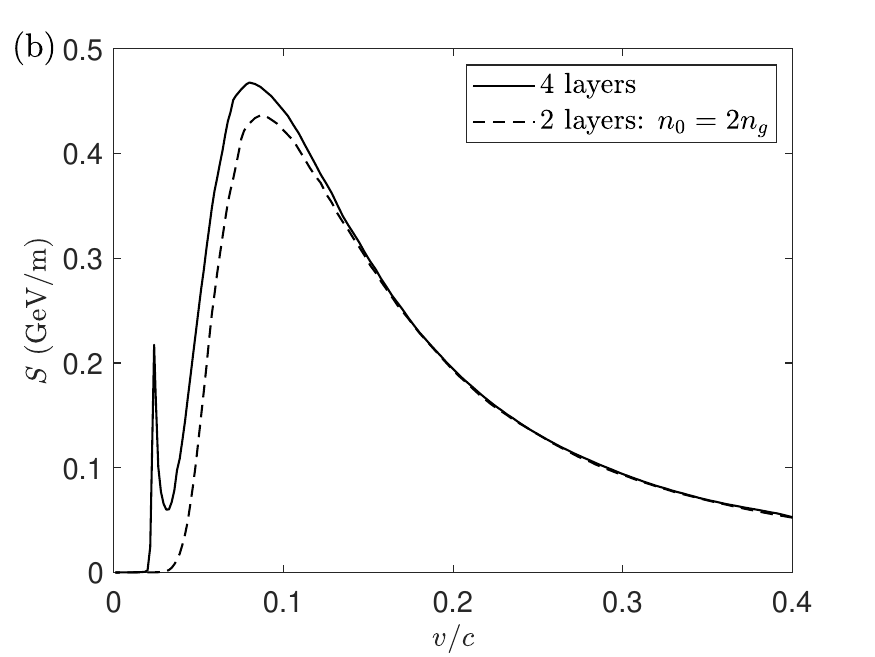}}
\centering
\caption{(a) Maximum longitudinal wakefield $W_x^{max}$ and (b) stopping power as a function of the driving velocity for the same configurations that Figure \ref{fig: 4_layer_xz_wakefields}, i.e. 4 layers at $z_4=-z_1=1.1$\,nm and $z_3=-z_2=0.9$\,nm vs 2 layers at $z_2=-z_1=1$\,nm with $n_0=2n_g$.}
\label{fig: 4_layer_max_wake}
\end{figure}

Finally, we are going to analyze what happens when the graphene layers are placed very close together. Thus, Figure \ref{fig: 4_layer_xz_wakefields} compares the wakefields in the  $\zeta z$-plane of 4 graphene layers (grouped in pairs) with the bi-layer configuration, considering $n_0=2n_g$. The results are very similar, with slight differences mainly due to the smaller aperture ($z_3-z_2=1.8$\,nm$<2$\,nm) in the case with 4 layers. Figure \ref{fig: 4_layer_max_wake} depicts the maximum wakefield $W_x^{max}$ and the stopping power as a function of the driving velocities for both configurations. It can be observed that they follow a similar trend, except that in the case with 4 layers a narrow peak is obtained at low velocities, similar to what occurs in double-walled CNTs \cite{Martin-Luna2024_DWCNTs_ResultsInPhysics}. Therefore, configurations with multiple grouped graphene layers can be used to create a system that behaves approximately like a single layer with higher density.

\section{Conclusions}\label{Conclusions}

The linearized hydrodynamic model has been used to describe the plasmonic excitations created by a point-like charge moving parallel to multilayer graphene that may be supported by an insulating substrate. In the proposed model, each layer can have a different surface density, allowing the analysis of a two-fluid model that makes distinction between $\sigma$ and $\pi$ electrons. We have formulated general expressions for the perturbed surface density, the induced potential and  both the longitudinal and transverse wakefields. A detailed analysis has been conducted on both the case of a single layer and a charged particle traveling between two layers. The dependence of the wakefields on different model parameters (driving velocity, position of the sheets, and surface density) has been studied, revealing that the intensity of the wakefields increases as the distance between the particle and the graphene layers decreases. It has also been observed that a witness beam cannot achieve simultaneous acceleration and focusing in the perpendicular plane (it focuses in one direction of the perpendicular plane and defocuses in the perpendicular direction). In the symmetric configuration of a particle traveling between two layers, the transverse wakefields are zero along the longitudinal axis, making the defocusing of a witness beam traveling slightly off-axis practically negligible along short distances, representing a potential candidate for particle acceleration purposes. Finally, the multilayer graphene has been analyzed, showing that if the layers are placed very close together, they behave approximately as a single layer with a surface density that is the sum of the densities of each individual layer. It has also been observed that adding more layers results in the appearance of new peaks in the representation of the maximum of the longitudinal wakefield $W_x^{max}$ as a function of the velocity, similar to what occurs in multi-walled CNTs. Ultimately, the findings presented in this paper could aid in designing specific types of multilayer graphene for potential applications like ion channeling, particle acceleration, and radiation emission. Future studies will delve deeper into the feasibility of these applications.

\section*{CRediT authorship contribution statement}
\textbf{Pablo Mart\'in-Luna:} Conceptualization, Methodology, Software, Validation,  Formal Analysis, Investigation, Writing - original draft, Writing - review \& editing. \textbf{Alexandre Bonatto:} Supervision, Writing - review \& editing. \textbf{Cristian Bontoiu:} Supervision, Writing - review \& editing. \textbf{Bifeng Lei:} Supervision, Writing - review \& editing. \textbf{Guoxing Xia:} Supervision, Writing - review \& editing. \textbf{Javier Resta-L\'opez:} Conceptualization, Methodology, Validation, Formal analysis, Investigation, Resources, Writing - original draft, Writing - review \& editing, Supervision, Project administration,  Funding acquisition.

\section*{Declaration of competing interest}
The authors declare that they have no known competing financial interests or personal relationships that could have appeared to influence the work reported in this paper.

\section*{Data availability}
Data will be made available on request.

\section*{Acknowledgments}
This work has been supported by Ministerio de Universidades (Gobierno de Espa\~{n}a) under grant agreement FPU20/04958, and the Generalitat Valenciana under grant agreement CIDEGENT/2019/058. 

\section*{References}

\begin{thebibliography}{10}
\expandafter\ifx\csname url\endcsname\relax
  \def\url#1{{\tt #1}}\fi
\expandafter\ifx\csname urlprefix\endcsname\relax\def\urlprefix{URL }\fi
\providecommand{\eprint}[2][]{\url{#2}}

\bibitem{Novoselov2004_graphene}
Novoselov K~S, Geim A~K, Morozov S~V, Jiang D, Zhang Y, Dubonos S~V, Grigorieva I~V and Firsov A~A 2004 {\em Science\/} {\bf 306} 666--669 \urlprefix\url{https://doi.org/10.1126/science.1102896}

\bibitem{LUO2013_plasmons_in_graphene_applications}
Luo X, Qiu T, Lu W and Ni Z 2013 {\em Materials Science and Engineering: R: Reports\/} {\bf 74} 351--376 ISSN 0927-796X \urlprefix\url{https://doi.org/10.1016/j.mser.2013.09.001}

\bibitem{Yan2008_vibrational_graphene_PhysRevB.77.125401}
Yan J~A, Ruan W~Y and Chou M~Y 2008 {\em Phys. Rev. B\/} {\bf 77}(12) 125401 \urlprefix\url{https://doi.org/10.1103/PhysRevB.77.125401}

\bibitem{wang2009vibrational_graphene}
Wang H, Wang Y, Cao X, Feng M and Lan G 2009 {\em Journal of Raman Spectroscopy: An International Journal for Original Work in all Aspects of Raman Spectroscopy, Including Higher Order Processes, and also Brillouin and Rayleigh Scattering\/} {\bf 40} 1791--1796 \urlprefix\url{https://doi.org/10.1002/jrs.2321}

\bibitem{cooper2012experimental_vibrational_graphene}
Cooper D~R, D’Anjou B, Ghattamaneni N, Harack B, Hilke M, Horth A, Majlis N, Massicotte M, Vandsburger L, Whiteway E {\em et~al.\/} 2012 {\em International Scholarly Research Notices\/} {\bf 2012} 501686 \urlprefix\url{https://doi.org/10.5402/2012/501686}

\bibitem{Nilsson2006_electronic_properties_graphene_PhysRevLett.97.266801}
Nilsson J, Neto A~H~C, Guinea F and Peres N~M~R 2006 {\em Phys. Rev. Lett.\/} {\bf 97}(26) 266801 \urlprefix\url{https://doi.org/10.1103/PhysRevLett.97.266801}

\bibitem{Novoselov2007_electronic_properties_graphene}
Novoselov K~S, Morozov S~V, Mohinddin T~M~G, Ponomarenko L~A, Elias D~C, Yang R, Barbolina I~I, Blake P, Booth T~J, Jiang D, Giesbers J, Hill E~W and Geim A~K 2007 {\em physica status solidi (b)\/} {\bf 244} 4106--4111 \urlprefix\url{https://doi.org/10.1002/pssb.200776208}

\bibitem{CastroNeto2009_electronic_properties_graphene_RevModPhys.81.109}
Castro~Neto A~H, Guinea F, Peres N~M~R, Novoselov K~S and Geim A~K 2009 {\em Rev. Mod. Phys.\/} {\bf 81}(1) 109--162 \urlprefix\url{https://doi.org/10.1103/RevModPhys.81.109}

\bibitem{Bonatto2023_effective_plasma_POP}
Bonatto A, Xia G, Apsimon O, Bontoiu C, Kukstas E, Rodin V {\em et~al.\/} 2023 {\em Physics of Plasmas\/} {\bf 30} 033105 \urlprefix\url{https://doi.org/10.1063/5.0134960}

\bibitem{Bontoiu2023_catapult}
Bonţoiu C, Apsimon O, Kukstas E, Rodin V, Yadav M, Welsch C {\em et~al.\/} 2023 {\em Scientific Reports\/} {\bf 13} 1330 \urlprefix\url{https://doi.org/10.1038/s41598-023-28617-w}

\bibitem{Martin-Luna2023_ExcitationWakefieldsSWCNT_NJP}
Martín-Luna P, Bonatto A, Bontoiu C, Xia G and Resta-López J 2023 {\em New Journal of Physics\/} {\bf 25} 123029 \urlprefix\url{https://doi.org/10.1088/1367-2630/ad127c}

\bibitem{Martin-Luna2024_DWCNTs_ResultsInPhysics}
Martín-Luna P, Bonatto A, Bontoiu C, Xia G and Resta-López J 2024 {\em Results in Physics\/} {\bf 60} 107698 ISSN 2211-3797 \urlprefix\url{https://doi.org/10.1016/j.rinp.2024.107698}

\bibitem{TajimaCavenago1987_XrayAccelerator_PhysRevLett.59.1440}
Tajima T and Cavenago M 1987 {\em Phys. Rev. Lett.\/} {\bf 59}(13) 1440--1443 \urlprefix\url{https://doi.org/10.1103/PhysRevLett.59.1440}

\bibitem{chen1987solid}
Chen P and Noble R~J 1987 A solid state accelerator {\em AIP Conf. Proc.\/} vol 156 pp 222--227 \urlprefix\url{https://doi.org/10.1063/1.36458}

\bibitem{chen1997crystal}
Chen P and Noble R~J 1997 Crystal channel collider: Ultra-high energy and luminosity in the next century {\em AIP Conf. Proc.\/} vol 398 pp 273--285 \urlprefix\url{https://doi.org/10.1063/1.53055}

\bibitem{Kinyanjui2012_EELS_graphene_monolayer}
Kinyanjui M~K, Kramberger C, Pichler T, Meyer J~C, Wachsmuth P, Benner G and Kaiser U 2012 {\em Europhysics Letters\/} {\bf 97} 57005 \urlprefix\url{https://doi.org/10.1209/0295-5075/97/57005}

\bibitem{Wachsmuth2013_EELS_monolayer_PhysRevB.88.075433}
Wachsmuth P, Hambach R, Kinyanjui M~K, Guzzo M, Benner G and Kaiser U 2013 {\em Phys. Rev. B\/} {\bf 88}(7) 075433 \urlprefix\url{https://doi.org/10.1103/PhysRevB.88.075433}

\bibitem{Eberlein2008_EELS_multilayer_PhysRevB.77.233406}
Eberlein T, Bangert U, Nair R~R, Jones R, Gass M, Bleloch A~L, Novoselov K~S, Geim A and Briddon P~R 2008 {\em Phys. Rev. B\/} {\bf 77}(23) 233406 \urlprefix\url{https://doi.org/10.1103/PhysRevB.77.233406}

\bibitem{FERMANIANKAMMERER2016_kinetic_model_layers}
{Fermanian Kammerer} C and Méhats F 2016 {\em Journal of Computational Physics\/} {\bf 327} 450--483 ISSN 0021-9991 \urlprefix\url{https://doi.org/10.1016/j.jcp.2016.09.010}

\bibitem{Allison2009_RPA_graphene_PhysRevB.80.195405}
Allison K~F, Borka D, Radovi\ifmmode~\acute{c}\else \'{c}\fi{} I, Had\ifmmode~\check{z}\else \v{z}\fi{}ievski L and Mi\ifmmode \check{s}\else \v{s}\fi{}kovi\ifmmode~\acute{c}\else \'{c}\fi{} Z~L 2009 {\em Phys. Rev. B\/} {\bf 80}(19) 195405 \urlprefix\url{https://doi.org/10.1103/PhysRevB.80.195405}

\bibitem{loche2018_molecular_dynamics_layers}
Loche P, Ayaz C, Schlaich A, Bonthuis D~J and Netz R~R 2018 {\em The journal of physical chemistry letters\/} {\bf 9} 6463--6468 \urlprefix\url{https://doi.org/10.1021/acs.jpclett.8b02473}

\bibitem{radovic2007hydrodynamic_model_layer}
Radovi{\'c} I, Had{\v{z}}ievski L, Bibi{\'c} N and Mi{\v{s}}kovi{\'c} Z~L 2007 {\em Physical Review A—Atomic, Molecular, and Optical Physics\/} {\bf 76} 042901 \urlprefix\url{https://doi.org/10.1103/PhysRevA.76.042901}

\bibitem{RADOVIC2010_hydrodynamic_model_layer_one_fluid}
Radović I and Borka D 2010 {\em Physics Letters A\/} {\bf 374} 1527--1533 ISSN 0375-9601 \urlprefix\url{https://doi.org/10.1016/j.physleta.2010.01.056}

\bibitem{RADOVIC2010_hydrodynamic_model_layer}
Radović I and Borka D 2010 {\em Nuclear Instruments and Methods in Physics Research Section B: Beam Interactions with Materials and Atoms\/} {\bf 268} 2649--2654 ISSN 0168-583X \urlprefix\url{https://doi.org/10.1016/j.nimb.2010.06.043}

\bibitem{LI2014_quantum_hydrodynamic_model_layer}
Li C~Z, Wang Y~N, Song Y~H and Mišković Z~L 2014 {\em Physics Letters A\/} {\bf 378} 1626--1631 ISSN 0375-9601 \urlprefix\url{https://doi.org/10.1016/j.physleta.2014.04.007}

\bibitem{BORKA2015_hydrodynamic_model_multilayer}
Borka D, Radović I and Vuković K 2015 {\em Nuclear Instruments and Methods in Physics Research Section B: Beam Interactions with Materials and Atoms\/} {\bf 347} 7--10 ISSN 0168-583X \urlprefix\url{https://doi.org/10.1016/j.nimb.2015.01.057}

\bibitem{Chaves2017_hydrodynamic_graphene_PhysRevB.96.195438}
Chaves A~J, Peres N~M~R, Smirnov G and Mortensen N~A 2017 {\em Phys. Rev. B\/} {\bf 96}(19) 195438 \urlprefix\url{https://doi.org/10.1103/PhysRevB.96.195438}

\bibitem{Hakimi2018_ionic_motion_PoP_10.1063/1.5016445}
Hakimi S, Nguyen T, Farinella D, Lau C~K, Wang H~Y, Taborek P {\em et~al.\/} 2018 {\em Physics of Plasmas\/} {\bf 25} 023112 \urlprefix\url{https://doi.org/10.1063/1.5016445}

\bibitem{Hakimi2020_ionic_motion_nanotube_doi:10.1142/S0217751X19430115}
Hakimi S, Zhang X, Lau C, Taborek P, Dollar F and Tajima T 2019 {\em International Journal of Modern Physics A\/} {\bf 34} 1943011 \urlprefix\url{https://doi.org/10.1142/S0217751X19430115}

\bibitem{Nejati2009_doi:10.1063/1.3077306}
Nejati M, Javaherian C, Shokri B and Jazi B 2009 {\em Physics of Plasmas\/} {\bf 16} 022108 \urlprefix\url{https://doi.org/10.1063/1.3077306}

\bibitem{Arista2001IonsPhysRevA.64.032901}
Arista N~R 2001 {\em Phys. Rev. A\/} {\bf 64}(3) 032901 \urlprefix\url{https://doi.org/10.1103/PhysRevA.64.032901}

\bibitem{Ostling1997_surface_density_PhysRevB.55.13980}
\"Ostling D, Tom\'anek D and Ros\'en A 1997 {\em Phys. Rev. B\/} {\bf 55}(20) 13980--13988 \urlprefix\url{https://doi.org/10.1103/PhysRevB.55.13980}

\bibitem{WangMiskovic2004_Hydro_theory_PhysRevA.69.022901}
Wang Y~N and Mi\ifmmode \check{s}\else \v{s}\fi{}kovi\ifmmode~\acute{c}\else \'{c}\fi{} Z~L 2004 {\em Phys. Rev. A\/} {\bf 69}(2) 022901 \urlprefix\url{https://doi.org/10.1103/PhysRevA.69.022901}

\bibitem{CHUNG2007_MWCNT_theory_RadPhysChem}
Chung S, Mowbray D, Mišković Z, Goodman F and Wang Y~N 2007 {\em Radiation Physics and Chemistry\/} {\bf 76} 524--528 proceedings of the 3rd International Conference on Elementary Processes in Atomic Systems \urlprefix\url{https://doi.org/10.1016/j.radphyschem.2005.09.020}

\end{thebibliography}
\providecommand{\noopsort}[1]{}\providecommand{\singleletter}[1]{#1}%
\providecommand{\newblock}{}

\end{document}